\documentclass[journal]{IEEEtran}
\usepackage{cite}
\usepackage{amsmath, amssymb}
\usepackage{amsfonts}
\usepackage{algorithmic}
\usepackage{graphicx}
\usepackage{textcomp}
\usepackage{physics}
\usepackage{unit}
\usepackage{bm}
\usepackage{multirow}
\usepackage{enumerate}
\usepackage{booktabs}
\usepackage{balance}
\usepackage{url}
\usepackage{threeparttable}
\usepackage{color}
\usepackage{xcolor}
\usepackage{siunitx}
\usepackage{colortbl}
\usepackage{tabularx}
\usepackage{pifont}
\usepackage[super]{nth}
\makeatletter
\let\MYcaption\@makecaption
\makeatother
\usepackage[font=footnotesize]{subcaption}

\makeatletter
\let\@makecaption\MYcaption
\makeatother
\usepackage{caption}

\def\BibTeX{{\rm B\kern-.05em{\sc i\kern-.025em b}\kern-.08em T\kern-.1667em\lower.7ex\hbox{E}\kern-.125emX}}

\begin{document}
    \title{Individual Identification Using Radar-Measured Respiratory and Heartbeat Features}

    \author{Haruto~Kobayashi, Yuji~Tanaka, and Takuya~Sakamoto
    \thanks{H.~Kobayashi and T.~Sakamoto are with the Department of Electrical Engineering, Graduate School of Engineering, Kyoto University, Kyoto 615-8510, Japan.}
    \thanks{Y.~Tanaka is with Graduate School of Engineering, Nagoya Institute of Technology, Nagoya 466-8555, Japan.}}

    \maketitle

    \begin{abstract}
        This study proposes a method for radar-based identification of individuals
        using a combination of their respiratory and heartbeat features. In the
        proposed method, the target individual's respiratory features are extracted
        using the modified raised-cosine-waveform model and their heartbeat features
        are extracted using the mel-frequency cepstral analysis technique. To
        identify a suitable combination of features and a classifier, we compare
        the performances of nine methods based on various combinations of three feature
        vectors with three classifiers. The accuracy of the proposed method in
        performing individual identification is evaluated using a 79-GHz
        millimeter-wave radar system with an antenna array in two experimental
        scenarios and we demonstrate the importance of use of the combination of
        the respiratory and heartbeat features in achieving accurate
        identification of individuals. The proposed method achieves accuracy of 96.33\%
        when applied to a five-day dataset of six participants and 99.39\% when applied
        to a public one-day dataset of thirty participants.
    \end{abstract}

    \section{Introduction}
    \label{sec:introduction} At present, with the ongoing advancements
    in Internet of Things (IoT) technology, techniques for sensing of various
    information types in real environments have been attracting attention. IoT technologies
    are easing the digitalization of society and enabling the construction of sensing
    networks for use in daily life~\cite{https://doi.org/10.1109/ACCESS.2023.3266647,
    https://doi.org/10.1109/JSEN.2021.3055618}. In particular, research into identification
    of individuals using sensors has drawn attention in a variety of applications.
    Currently, facial recognition using cameras is primarily being researched and
    commercialized~\cite{https://doi.org/10.1109/CVPR.2015.7298682, https://doi.org/10.1109/TPAMI.2019.2961900}.
    Although facial recognition can achieve high-precision identification, it also
    raises privacy concerns and is unsuitable for sensing applications in private
    spaces. Additionally, external factors, such as low lighting, can lead to reduced
    recognition accuracy.

    Rather than use cameras, contact sensor-based methods for individual
    identification that sense fingerprints~\cite{https://doi.org/10.1109/IACC.2016.69},
    veins~\cite{https://doi.org/10.1109/TIP.2009.2023153}, and
    electrocardiograms (ECGs)~\cite{https://doi.org/10.1109/TBIOM.2019.2947434, https://doi.org/10.1109/TIM.2022.3199260,
    https://doi.org/10.1109/TIM.2024.3368481}
    have been reported. Furthermore, individual identification techniques using
    respiratory sounds measured with microphones have also been reported~\cite{https://doi.org/10.1109/INFOCOM53939.2023.10229037,
    https://doi.org/10.1145/3081333.3081355}. Han \emph{et al.}~\cite{https://doi.org/10.1109/INFOCOM53939.2023.10229037}
    proposed an individual identification method that used respiratory sounds
    acquired from bone conduction microphones embedded in earphones and achieved
    identification accuracy of 95.17\% for 20 participants. Although these
    studies focused on use of contact sensors to obtain the physiological
    signals required for individual identification, it can be difficult to attach
    the sensors or place the required sensors close to the body, and this factor
    will also hinder the identification of uncooperative individuals.

    Recently, radar has been used widely for measurement of human bodies because
    both limb movements and physiological signals can be acquired in a noncontact
    manner and privacy concerns are reduced when compared with use of camera-based
    sensors~\cite{https://doi.org/10.1109/JSEN.2020.3036039, https://doi.org/10.1109/TBCAS.2022.3226290,
    https://doi.org/10.1109/JSEN.2020.3023486}. Radar techniques have been used
    for respiration measurement ~\cite{https://doi.org/10.1109/JSEN.2021.3117707,
    https://doi.org/10.1109/TMC.2015.2504935, https://doi.org/10.1109/JTEHM.2014.2365776},
    heartbeat~\cite{https://doi.org/10.1109/LSENS.2023.3322287, https://doi.org/10.1109/ACCESS.2019.2912956,
    itsuki, https://doi.org/10.1109/TBME.2015.2470077,https://doi.org/10.1109/JSEN.2022.3148003,https://doi.org/10.1109/JSEN.2019.2950635}
    and also for individual identification applications~\cite{https://doi.org/10.1109/COMST.2024.3373397,
    https://doi.org/10.1109/JPROC.2023.3244362}.

    Limb motions such as walking and sitting movements are often used to
    identify individuals, e.g., in individual identification of walking and sitting
    persons using Wi-Fi signals~\cite{https://doi.org/10.1109/GLOCOM.2016.7841847,
    https://doi.org/10.1109/IPSN.2016.7460727, https://doi.org/10.1145/2942358.2942373,
    https://doi.org/10.1109/JIOT.2020.3037945, https://doi.org/10.1109/INFOCOM41043.2020.9155258}
    and radar techniques~\cite{https://doi.org/10.1049/iet-rsn.2017.0511, https://doi.org/10.1109/TGRS.2018.2816812,
    https://arxiv.org/abs/2008.02182, https://doi.org/10.1109/LSENS.2020.2975219,
    https://doi.org/10.1049/iet-rsn.2019.0618, https://doi.org/10.1109/JSEN.2023.3299558,
    https://doi.org/10.1109/JSEN.2020.3032960, https://doi.org/10.1109/JSEN.2022.3165207,
    https://doi.org/10.1109/TGRS.2020.3019915, https://doi.org/10.1109/TCSVT.2022.3161515,
    https://doi.org/10.1109/JIOT.2019.2929833, https://doi.org/10.1109/TGRS.2021.3073664}.
    To enable individual identification of a person that is remaining still, Wi-Fi
    signals and radar have also been used to measure their physiological signals~\cite{https://doi.org/10.1109/JIOT.2023.3275099,
    https://doi.org/10.1109/JIOT.2024.3358548, https://doi.org/10.1109/JIOT.2024.3395287,
    https://doi.org/10.1109/JETCAS.2018.2818181, https://doi.org/10.1109/JIOT.2022.3196143,
    https://doi.org/10.1109/ACCESS.2023.3328641, https://doi.org/10.1109/SECON55815.2022.9918606,
    https://doi.org/10.1109/JSEN.2024.3353256, https://doi.org/10.1145/3117811.3117839}.
    For example, Rahman \emph{et al.}~\cite{https://doi.org/10.1109/JETCAS.2018.2818181}
    extracted features from radar-measured respiration characteristics and
    identified individuals using a combination of the $k$-nearest neighbor ($k$-NN)
    algorithm and majority voting. Although respiration measured via radar is useful
    for individual identification, respiration is also known to have both
    voluntary and involuntary aspects~\cite{https://doi.org/10.1242/jeb.100.1.93},
    which raises questions about its effectiveness for use in long-term identification.
    In contrast, Lin \emph{et al.}~\cite{https://doi.org/10.1145/3117811.3117839}
    conducted individual identification using heartbeat features that were
    measured using a pair of radar systems, with accuracy of 98.61\% being
    achieved for 78 participants.

    In this paper, we propose an individual identification method that uses both
    respiratory and heartbeat features for the first time. A preprint of this
    manuscript has been posted online \cite{preprint}. The contributions of this
    paper are listed as follows:
    \begin{itemize}
        \item We propose a method to extract respiratory features in addition to
            extraction of respiratory intervals, and confirm the effectiveness
            of using these features to perform individual identification through
            experiments.

        \item We propose a method to extract features of heartbeat harmonic components
            and demonstrate the effectiveness of use of these features in individual
            identification.

        \item Through a series of experiments conducted over five days on six participants,
            we evaluate the long-term effectiveness of the proposed method. We also
            verify the method's accuracy when using the respiratory and heartbeat
            features separately and discuss the advantages of combining these features.
            Additionally, we validate the identification accuracy for larger groups
            using a public dataset with 30 participants.
    \end{itemize}

    \section{Respiratory Features for Individual Identification}
    \label{sec:resp}
    \subsection{Waveform Feature Extraction Method}
    \label{subsec:MRCW} If the radar system receives a single dominant echo from
    a human body, the body's displacement can be estimated to be
    \begin{align}
        d(t) = \frac{\lambda}{4\pi}\mathrm{unwrap}\qty{\angle s(t)}, \label{eq:phase_demodulation}
    \end{align}
    where $\lambda$ is the wavelength, $s(t)$ is the signal reflected from the target,
    $\mathrm{unwrap}\{\cdot\}$ is the phase unwrapping operator, and $\angle$ represents
    the phase of a complex number. When the target person is stationary, $d(t)$ mainly
    contains the displacement components that result from the person's respiration
    and heartbeat.

\begin{figure}[tb!]
        \centering
        \includegraphics[width=0.7\linewidth]{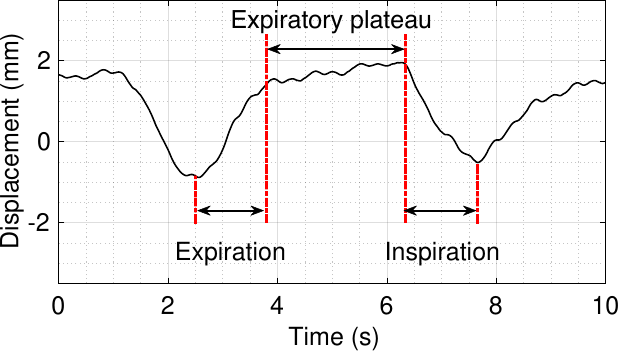}
        \caption{Example of respiratory displacement when measured using a radar
        system.}
        \label{fig:resp_ex}
    \end{figure}

    The respiratory displacement caused by normal breathing can be expressed as a
    semi-periodic function of time with three phases: the expiratory phase, the expiratory
    plateau, and the inspiratory phase~\cite{https://doi.org/10.1093/bjaed/mkw062}.
    Fig.~\ref{fig:resp_ex} shows an example of a radar-measured displacement
    waveform that highlights the three phases described above. To extract individual
    features related to the three respiratory phases, we use the modified raised-cosine-waveform
    (MRCW) model proposed by Hsieh \emph{et al.}~\cite{https://doi.org/10.1109/ISCAS.2013.6572037}.
    In the MRCW model, in addition to the respiratory period and amplitude, the
    user can also specify the time duration ratio of the expiratory, expiratory plateau,
    and inspiratory phases.

    The MRCW model for one cycle $T=1/f$ can be expressed using six parameters
    as follows:
    \begin{align}
        \begin{split}&d_{\mathrm{MRCW}}(t, A, f, \beta_{1}, \beta_{2}, D)= \\&\hspace*{-5pt}\left\{\begin{aligned}&A&&\hspace*{-10pt}\qty(\displaystyle \frac{T}{2}- T_{\mathrm{a1}}<t\leq\frac{T}{2}+ T_{\mathrm{a2}})&\\&A\cos\qty{\displaystyle \frac{2\pi f}{\beta_1}\qty(\qty|t-\frac{T}{2}|-T_\mathrm{a1})}&&\hspace*{-10pt}\qty(\displaystyle T_{\mathrm{b1}}<t\leq\frac{T}{2}- T_{\mathrm{a1}})&\\&A\cos\qty{\displaystyle \frac{2\pi f}{\beta_2}\qty(\qty|t-\frac{T}{2}|-T_\mathrm{a2})}&&\hspace*{-10pt}\qty(\displaystyle \frac{T}{2}+T_{\mathrm{a2}}<t\leq T - T_{\mathrm{a1}})&\\&-A&&\hspace*{-3pt}\text{otherwise,}&\end{aligned}\right.\end{split} \label{eq:MRCW}
    \end{align}
    where $f$ is the respiratory frequency, $A$ is the amplitude, $\beta_{1}$ and
    $\beta_{2}$ are the roll-off rates, and $D$ is the duty ratio. Note that $\beta
    _{1}$, $\beta_{2}$, and $D$ correspond to the temporal ratios of the expiration,
    inspiration, and expiratory plateau phases, respectively. In addition,
    $T_{\mathrm{a1}}, T_{\mathrm{a2}}, T_{\mathrm{b1}},$ and $T_{\mathrm{b2}}$
    in \eqref{eq:MRCW} are determined as follows:
    \begin{align}
         & T_{\mathrm{a1}}= \frac{D(1-\beta_{1})}{2f},\, T_{\mathrm{b1}}= \frac{(1-D)(1-\beta_{1})}{2f}, \\
         & T_{\mathrm{a2}}= \frac{D(1-\beta_{2})}{2f},\, T_{\mathrm{b2}}= \frac{(1-D)(1-\beta_{2})}{2f}.
    \end{align}

\begin{figure}[tb!]
        \centering
        \begin{minipage}{0.49\linewidth}
            \centering
            \includegraphics[width=\linewidth]{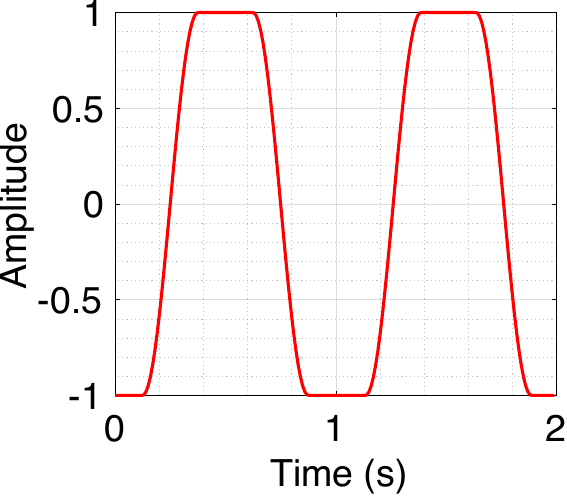}
            \subcaption{}
        \end{minipage}
        \begin{minipage}{0.49\linewidth}
            \centering
            \includegraphics[width=\linewidth]{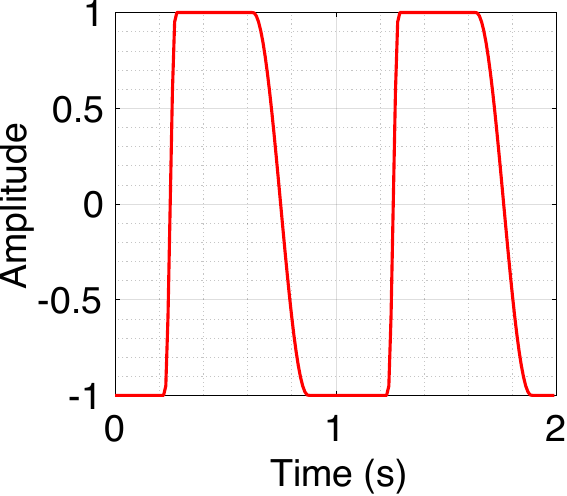}
            \subcaption{}
        \end{minipage}\\
        \vspace*{10pt}
        \begin{minipage}{0.49\linewidth}
            \centering
            \includegraphics[width=\linewidth]{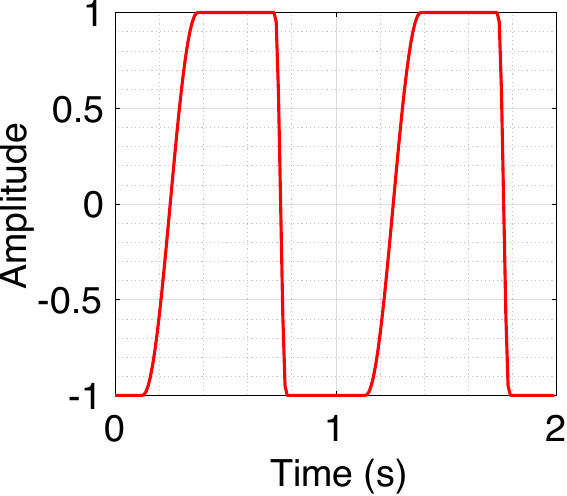}
            \subcaption{}
        \end{minipage}
        \begin{minipage}{0.49\linewidth}
            \centering
            \includegraphics[width=\linewidth]{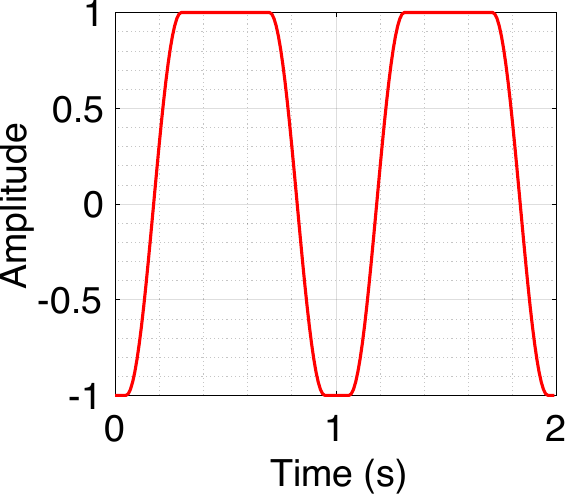}
            \subcaption{}
        \end{minipage}
        \caption{Examples of the respiratory displacement MRCW model with
        parameters (a) $(\beta_{1},\beta_{2},D)=(0.5, 0.5, 0.5)$, (b) $(\beta_{1}
        ,\beta_{2},D)=(0.1, 0.5, 0.5)$, (c)
        $(\beta_{1},\beta_{2},D)=(0.5, 0.1, 0.5)$, and (d) $(\beta_{1},\beta_{2},
        D)=(0.5, 0.5, 0.8)$.}
        \label{fig:mrcw}
    \end{figure}

    Example MRCW models are shown in Fig.~\ref{fig:mrcw} for $f=1$~Hz and $A=1$.
    As the figure shows, the waveform for the MRCW model is dependent on the
    parameters $\qty(f, \beta_{1},\beta_{2},D)$. Hereafter, the parameters
    $\bm{p}=\qty(f, \beta_{1},\beta_{2},D)^{\mathrm{T}}$ are used as the
    respiratory features (a superscript $\mathrm{T}$ signifies the transpose operator).

    To determine the MRCW model parameters from the measured signal, the following
    operation is performed:
    \begin{align}
        \begin{split}\qty(\hat{\tau}(t), \hat{A}(t),\hat{\bm{p}}(t))&= \\ \underset{(\tau, A,\bm{p})}{\operatorname{argmin}}\int_{0}^{T_0}&\left|d(t')w(t'-t) - d_{\mathrm{MRCW}}\qty(t'-\tau, A,\bm{p})\right|^{2}\,\dd t' , \label{eq:est}\end{split} \\
        \qquad\hat{d}_{\mathrm{MRCW}}(t)                                                                                                                                                                                                               & = d_{\mathrm{MRCW}}\qty(t-\hat{\tau}(t), \hat{A}(t),\hat{\bm{p}}(t)),
    \end{align}
    where $T_{0}$ is the measurement time length, $\tau$ is the time shift and $\hat
    {d}_{\mathrm{MRCW}}(t)$ is the MRCW model that provides the best approximation
    of $d(t)$. In addition, $w(t)$ is a rectangular window with a length of 8 s.
    The parameters estimated in \eqref{eq:est} are defined here as the estimated
    waveform features $\hat{\bm{p}}(t)=(\hat{f}(t), \hat{\beta}_{1}(t),\hat{\beta}
    _{2}(t),\hat{D}(t))^{\mathrm{T}}$ at a time $t$.

\begin{figure}[tb!]
        \centering
        \includegraphics[width=0.8\linewidth]{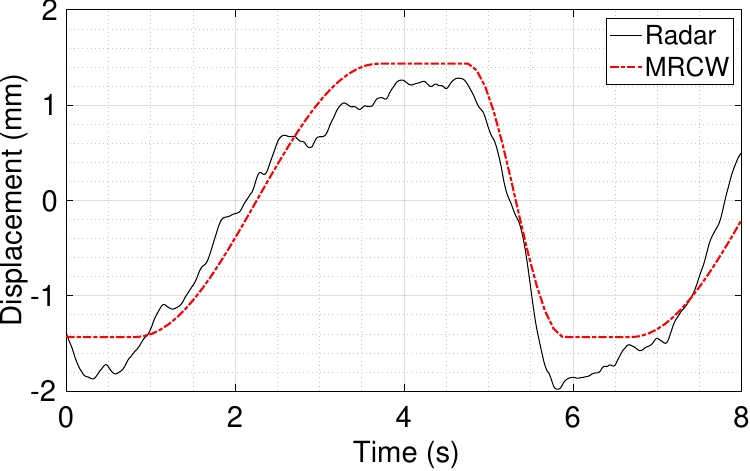}
        \caption{Displacement waveform measured using radar (black solid line)
        and calculated using the MRCW model (red dashed line).}
        \label{fig:disp_MRCW}
    \end{figure}

    The measured displacement waveform $d(t)$ and the estimated waveform from the
    MRCW model $\hat{d}_{\mathrm{MRCW}}(t)$ are shown in Fig.~\ref{fig:disp_MRCW}.
    This comparison indicates that $\hat{d}_{\mathrm{MRCW}}(t)$ approximates $d(t
    )$.
    Note that the MRCW model parameters in Fig.~\ref{fig:disp_MRCW} are $\hat{f}=
    0.4$~Hz and $(\hat{\beta}_{1}, \hat{\beta}_{2},\hat{D})=(0.9, 0.6, 0.9)$.

    \subsection{Features of the Respiratory Expiratory Plateau}
    Although the MRCW model can approximate the respiratory displacement, the
    expiratory plateau waveform is modeled as a constant straight line, which is
    inaccurate, as illustrated in Fig.~\ref{fig:disp_MRCW}. In this study, the displacement
    waveform during the expiratory plateau time interval is used as an additional
    feature.

\begin{figure}[tb!]
        \centering
        \includegraphics[width=0.8\linewidth]{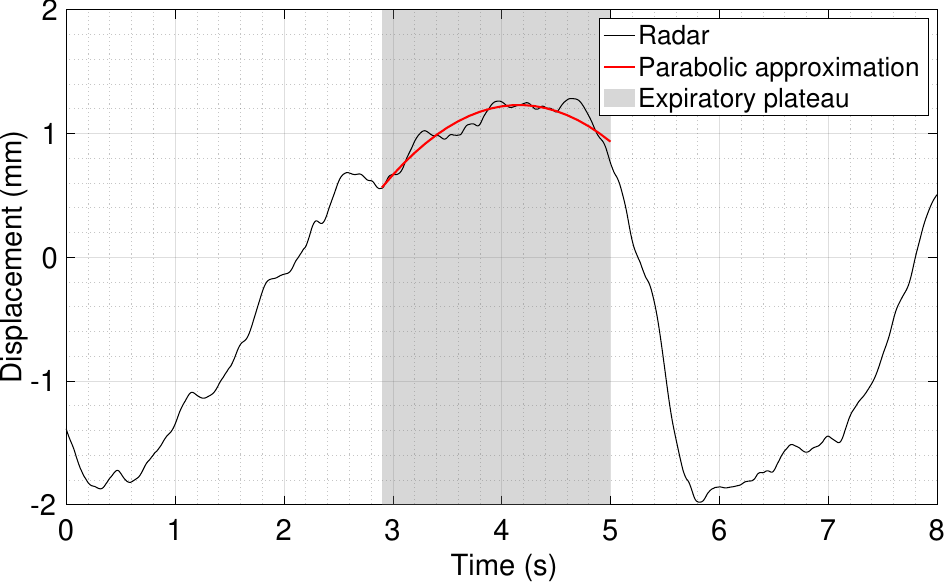}
        \caption{Displacement waveform measured using radar (black solid line)
        and calculated using the parabolic model (red solid line).}
        \label{fig:2ndcurve}
    \end{figure}

    For $t$, parabolic fitting is performed using the parabolic curve
    $d_{\mathrm{par}}(t;c_{0},c_{1},c_{2})=c_{2}t^{2}+ c_{1}t + c_{0}$ as shown:
    \begin{align}
        \begin{aligned}&\qty(\hat{c}_{0}(t), \hat{c}_{1}(t), \hat{c}_{2}(t))= \\&\quad \underset{(c_0, c_1, c_2)}{\operatorname{argmin}}\int_{0}^{T_0}\qty|w_{\mathrm{E}}^{(t)}(t')\qty{d(t') - d_\mathrm{par}(t';c_0,c_1,c_2)}|^{2}\,\dd t',\end{aligned} \label{eq:2ndcurve}
    \end{align}
    where $w_{\mathrm{E}}^{(t)}(t')$ is a window function that is defined as follows:
    \begin{align}
        w_{\mathrm{E}}^{(t)}(t') = \begin{cases}w(t'-t)&\qty(\hat{d}_{\mathrm{MRCW}}(t')\geq\gamma\hat{A}(t)), \\ 0&(\text{otherwise}),\end{cases}
    \end{align}
    where $\gamma=0.6$ is a threshold that was set empirically. In this paper,
    $\hat{c}_{2}(t)$ is used as a feature that represents the expiratory plateau's
    shape. Fig.~\ref{fig:2ndcurve} shows the time interval for the expiratory
    plateau (gray area) and the corresponding estimated parabolic curve (red solid
    line).

    \subsection{Features of the Respiratory Statistics}
    In Section~\ref{subsec:MRCW}, a method to estimate the respiratory features
    at each instant $t$ was presented. However, respiration includes both
    voluntary and involuntary components that result in temporal fluctuations in
    the respiratory feature. These temporal variations can have an adverse effect
    on the accuracy of individual identification. To mitigate this variability, we
    extract the statistical properties of the respiratory features.

    First, the MRCW parameters
    $(\hat{f}(t),\hat{\beta}_{1}(t), \hat{\beta}_{2}(t), \hat{D}(t))$ are
    transformed into a six-dimensional vector $\bm{q}(t)$, as follows:
    \begin{align}
        \bm{q}(t) = \begin{bmatrix}q_{1}\\q_{2}\\ q_{3}\\q_{4}\\ q_{5}\\ q_{6}\end{bmatrix}=\begin{bmatrix}\hat{f}(t)\\ \hat{D}(t)\\ \hat{\beta}_{1}(t)+\hat{\beta}_{2}(t)\\ |\hat{\beta}_{1}(t)-\hat{\beta}_{2}(t)| \\ \qty{\hat{\beta}_1(t)+\hat{\beta}_2(t)}/\hat{f}(t) \\ \hat{c}_{2}(t)\end{bmatrix}. \label{eq:feature}
    \end{align}
    In \eqref{eq:feature}, $q_{1}$ and $q_{2}$ represent the respiratory
    frequency and the expiratory plateau duration, respectively. Furthermore, $q_{3}$
    and $q_{4}$ represent the sum and the difference, respectively, of the inspiratory
    and expiratory rates. Additionally, $q_{5}$ represents the ratio of the inspiratory
    and expiratory rates to the sum of the inspiratory and expiratory rates relative
    to the respiratory frequency. Finally, $q_{6}$ is a coefficient of the parabolic
    curve estimated using \eqref{eq:2ndcurve}.

    Next, statistics are extracted for each element of the six-dimensional feature
    vector $\bm{q}(t)$ and transformed into a 24-dimensional feature vector $\bm{r}
    _{\mathrm{resp}}$ as $\bm{r}_{\mathrm{resp}}=[\bm{r}_{1},\bm{r}_{2},\ldots,\bm
    {r}_{6}]^{\mathrm{T}}\in\mathbb{R}^{24\times 1}$, where
    $\bm{r}_{n}= [\mu_{n},\sigma_{n},\gamma_{n},\kappa_{n}]\in\mathbb{R}^{1\times
    4}$
    is defined for $n=1,2,\ldots,6$; here, $\mu_{n}$, $\sigma_{n}$, $\gamma_{n}$,
    and $\kappa_{n}$ represent the mean, the standard deviation, the skewness,
    and the kurtosis of $q_{n}$, respectively.

    \section{Heartbeat Features for Individual Identification}
    \label{sec:heart_exp}
    \subsection{High-Frequency Enhancement Using Differentiation}
    \label{subsec:diff} In this section, we propose a method to extract the
    features of the heartbeat component required for individual identification. Because
    the fundamental frequency component of the heartbeat is often masked by harmonic
    components of the respiration, it has proven to be advantageous to use the
    higher harmonics of the heartbeat components~\cite{https://doi.org/10.1109/LSENS.2023.3322287,
    https://doi.org/10.1109/JSEN.2022.3148003,https://doi.org/10.1109/JSEN.2019.2950635}.
    Therefore, our proposed method emphasizes the high-frequency components by
    using second-order differentiation, as in our previous work~\cite{itsuki}.
    Because the second-order differentiation $s''(t)=(\dd^{2}/\dd t^{2})s(t)$ is
    expressed as $-\omega^{2}S(\omega)$ in the frequency domain, the high-frequency
    components, including the harmonics of the heartbeat, are emphasized; here, $\omega$
    is the angular frequency and $S(\omega)$ represents the Fourier transform of
    $s(t)$.

    \subsection{Feature Extraction Using Mel-Frequency Cepstrum}
    Next, we use the mel-frequency cepstral coefficients (MFCCs) to generate
    features. Historically, MFCCs have been used to convert the short-term
    spectra of speech signals into feature vectors~\cite{https://doi.org/10.1109/JPROC.2013.2251592}.
    Because radar-measured heartbeat signals are periodic within a short time,
    in exactly the same manner as speech signals, use of MFCCs to generate heartbeat
    features is appropriate when using radar~\cite{https://doi.org/10.1109/JSEN.2024.3353256}.
    In~\cite{https://doi.org/10.1109/JSEN.2024.3353256}, MFCCs were applied to real
    signals that were estimated from the phases of reflected signals. However, our
    previous work~\cite{itsuki} has shown that signal processing using complex reflected
    signals directly improves the heartbeat estimation accuracy. Therefore, in this
    section, we propose a method to extract heartbeat features by applying MFCCs
    to complex signals. The calculation of the MFCCs is performed using the following
    steps.

    \subsubsection{Short-time Fourier transform}
    For the differentiated signal $s''(t)$, the short-time Fourier transform (STFT)
    is applied as follows:
    \begin{align}
        S(t, f) = \int_{-\infty}^{\infty}s''(\tau) w_{\mathrm{STFT}}(\tau - t)\exp\left(-\jj2\pi f \tau\right)\dd\tau, \label{eq:STFT}
    \end{align}
    where $w_{\mathrm{STFT}}(t)$ is a rectangular window function with a width of
    $T_{\mathrm{STFT}}$. This window width $T_{\mathrm{STFT}}$ should be greater
    than the heartbeat interval and we set $T_{\mathrm{STFT}}=2.0$ s empirically.

    \subsubsection{Mel filters}
\begin{figure}[tb!]
        \centering
        \includegraphics[width=\linewidth]{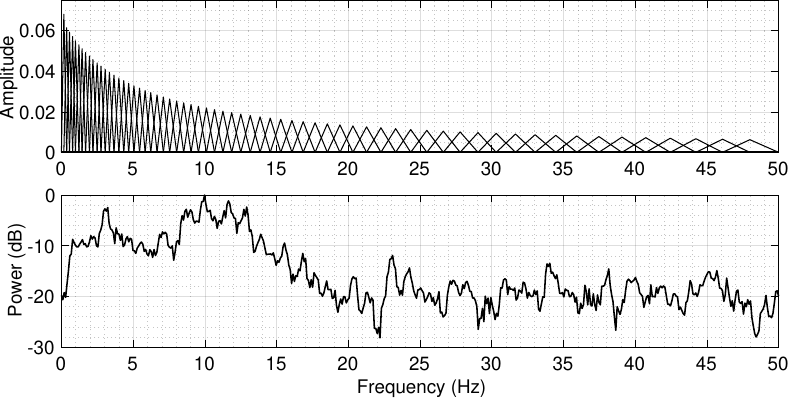}
        \caption{Examples of the mel filters and the power spectrum for $s''(t)$.}
        \label{fig:mel_filter}
    \end{figure}

    Next, we apply a mel filter bank to $S(t,f)$ to calculate the mel-frequency spectrum.
    The $\ell$th mel filter $H_{\ell}(f)$ is expressed as:
    \begin{align}
        H_{\ell}(f) & = \begin{cases}\vspace{2mm}\displaystyle \frac{2(f - f_{\ell})}{(f_{\ell+1} - f_{\ell})(f_{\ell+2} - f_{\ell})}&\displaystyle (f_{\ell}\leq f < f_{\ell+1}), \\ \vspace{2mm}\displaystyle \frac{2(f_{\ell+2} - f)}{(f_{\ell+2} - f_{\ell+1})(f_{\ell+2} - f_{\ell})}&\displaystyle (f_{\ell+1}\leq f < f_{\ell+2}), \\ 0&(\text{otherwise})\end{cases}
    \end{align}
    for $\ell=0,1,\ldots,L-1$, where
    \begin{align}
        f_{\ell} & = \tilde{f}\qty{\exp\qty(\frac{m_{\ell}}{\tilde{m}})-1},
    \end{align}
    is the center frequency of the $\ell$th mel filter corresponding to the
    $\ell$th mel scale. Here, the mel scale $m_{\ell}$ is defined as
    \begin{align}
        m_{\ell}= \tilde{m}\frac{\ell}{L+1}\log\left(1 + \frac{f_{\mathrm{s}}}{2\tilde{f}}\right), \label{eq:mel}
    \end{align}
    for $\ell = 0,1,\ldots,L+1$, where $f_{\mathrm{s}}$ is the sampling
    frequency and $\tilde{m}$ is defined as:
    \begin{align}
        \tilde{m} & = \frac{f'}{\log\qty(\displaystyle f'/\tilde{f} + 1)}.
    \end{align}
    An example of a mel filter bank and the power spectrum of $s''(t)$ are shown
    in Fig.~\ref{fig:mel_filter} for the case where $\tilde{f}=5.0$~Hz, $f'=1.0$~kHz,
    and $L=64$, where these parameters were set following the convention in \cite{https://doi.org/10.1109/JSEN.2024.3353256}.

    In this study, we apply the mel filters to a complex radar signal, where the
    mel filters are applied to the spectrum for both positive and negative
    frequencies as follows:
    \begin{align}
        S_{\ell}  & = \int_{0}^{T_0}\int_{0}^{f_\mathrm{s}/2}|S(t, f)| H_{\ell}(f) \, \dd f\dd t, \label{eq:incoherent_int1}   \\
        S_{-\ell} & = \int_{0}^{T_0}\int_{-f_\mathrm{s}/2}^{0}|S(t, f)| H_{\ell}(-f) \, \dd f \dd t,\label{eq:incoherent_int2}
    \end{align}
    for $0\leq \ell \leq L-1$, where $T_{0}$ is the measurement time. Note that in
    this case, $S_{+0}$ and $S_{-0}$ are regarded as different variables.

    \subsubsection{Discrete cosine transform}
    Next, we obtain the MFCC $C_{k}$ by applying a discrete cosine transform (DCT)
    to $\log S_{\ell}$ as shown here:
    \begin{align}
        C_{k}=  & \frac{2}{L+\delta_{k,0}}\sum_{\ell=0}^{L-1}\log S_{\ell}\cos\qty{\frac{(2\ell+1)k\pi}{K}},    \\
        C_{-k}= & \frac{2}{L+\delta_{-k,0}}\sum_{\ell=0}^{L-1}\log S_{-\ell}\cos\qty{\frac{-(2\ell+1)k\pi}{K}},
    \end{align}
    for $k=0,1,\ldots,K-1$, where $\delta_{i,j}$ is the Kronecker delta, $K$ is
    the cepstrum dimension, and we set $K=64$. Note that $C_{+0}$ and $C_{-0}$
    are also regarded as different variables here.

    Finally, from the $2K$-dimensional MFCC $C_{k}$, the lower order
    coefficients up to $2K'$ dimensions are extracted, where the condition $0< K'
    <K$ holds; the feature vector of the heartbeat
    $\bm{r}_{\mathrm{hb}}=[C_{-(K'-1)},C_{-(K'-2)},\ldots,C_{-0},C_{+0},\ldots,C_{K'-1}
    ]^{\mathrm{T}}$
    is defined and is used to perform individual identification in the following
    section, where we set $K'=24$ empirically.

    \section{Performance Evaluation of the Proposed Method}
    \label{sec:exp} In this section, we evaluate the individual identification
    performance of the proposed method via comparison with conventional methods.
    To identify an effective set of features, we compare the performance metrics
    obtained when the proposed feature vectors are used separately and together.
    We compare three cases: (1) use of the respiratory feature vector
    $\bm{r}_{\mathrm{resp}}$ only; (2) use of the heartbeat feature vector $\bm{r}
    _{\mathrm{hb}}$ only; and (3) use of both vectors as
    $\bm{r}_{\mathrm{prop}}=[\bm{r}_{\mathrm{resp}}^{\mathrm{T}}, \bm{r}_{\mathrm{hb}}
    ^{\mathrm{T}}]^{\mathrm{T}}$. We also compare three classifiers: the support
    vector machine (SVM), $k$-nearest neighbors ($k$-NN), and multilayer
    perceptron (MLP) classifiers. For these three classifiers, we use the three
    feature vector types, i.e., $\bm{r}_{\mathrm{resp}}$, $\bm{r}_{\mathrm{hb}}$,
    and $\bm{r}_{\mathrm{prop}}$, to compare a total of nine different
    approaches. Table~\ref{tbl:Sakamoto} summarizes the definitions of nine methods
    for comparison; for example, method A1 indicates the identification process when
    using $\bm{r}_{\mathrm{resp}}$ and the SVM.

    \begin{table}[tb!]
        \centering
        \caption{DEFINITIONS OF METHODS FOR INDIVIDUAL IDENTIFICATION}
        \begin{tabular}{ccccc}
            \toprule \textbf{Method}     & \textbf{Signal}                                                                                                          & \textbf{Vital sign}                                                                     & \textbf{Feature}                          & \textbf{Classifier} \\
            \midrule Method A1           & \multirow{3}{*}{$\angle s(t)$}                                                                                           & \multirow{3}{*}{Respiration}                                                            & \multirow{3}{*}{$\bm{r}_{\mathrm{resp}}$} & SVM                 \\
            Method A2                    &                                                                                                                          &                                                                                         &                                           & $k$-NN              \\
            Method A3                    &                                                                                                                          &                                                                                         &                                           & MLP                 \\
            \cmidrule(lr){1-5} Method B1 & \multirow{3}{*}{$\frac{\mathrm{d}^{2}}{\mathrm{d}t^{2}}s(t)$}                                                            & \multirow{3}{*}{Heartbeat}                                                              & \multirow{3}{*}{$\bm{r}_{\mathrm{hb}}$}   & SVM                 \\
            Method B2                    &                                                                                                                          &                                                                                         &                                           & $k$-NN              \\
            Method B3                    &                                                                                                                          &                                                                                         &                                           & MLP                 \\
            \cmidrule(lr){1-5} Method C1 & \multirow{3}{*}{\begin{tabular}[c]{@{}c@{}}$\angle s(t)$ \&\\ $\frac{\mathrm{d}^{2}}{\mathrm{d}t^{2}}s(t)$\end{tabular}} & \multirow{3}{*}{\begin{tabular}[c]{@{}c@{}}Respiration \\ \& \\ heartbeat\end{tabular}} & \multirow{3}{*}{$\bm{r}_{\mathrm{prop}}$} & SVM                 \\
            Method C2                    &                                                                                                                          &                                                                                         &                                           & $k$-NN              \\
            Method C3                    &                                                                                                                          &                                                                                         &                                           & MLP                 \\
            \bottomrule
        \end{tabular}
        \label{tbl:Sakamoto}
    \end{table}

    \subsection{Experiment 1: Five-Day Measurement of Six Participants}
    \label{subsec:exp1} 
\begin{figure}[tb!]
        \centering
        \begin{minipage}{0.99\linewidth}
            \centering
            \includegraphics[width=0.7\linewidth]{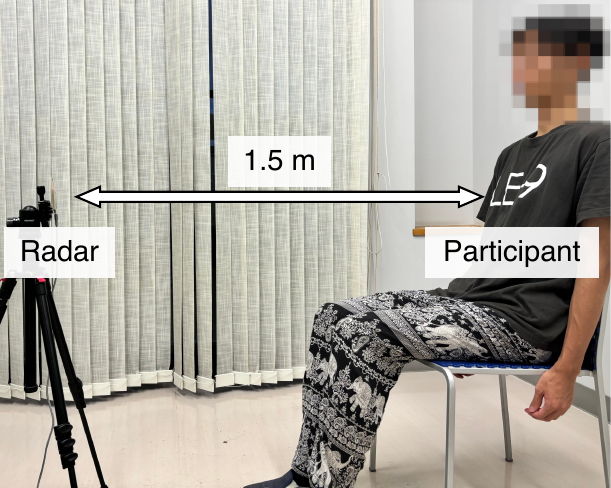}
            \subcaption{} \label{fig:scene}
        \end{minipage}
        \begin{minipage}{0.99\linewidth}
            \centering
            \includegraphics[width=\linewidth]{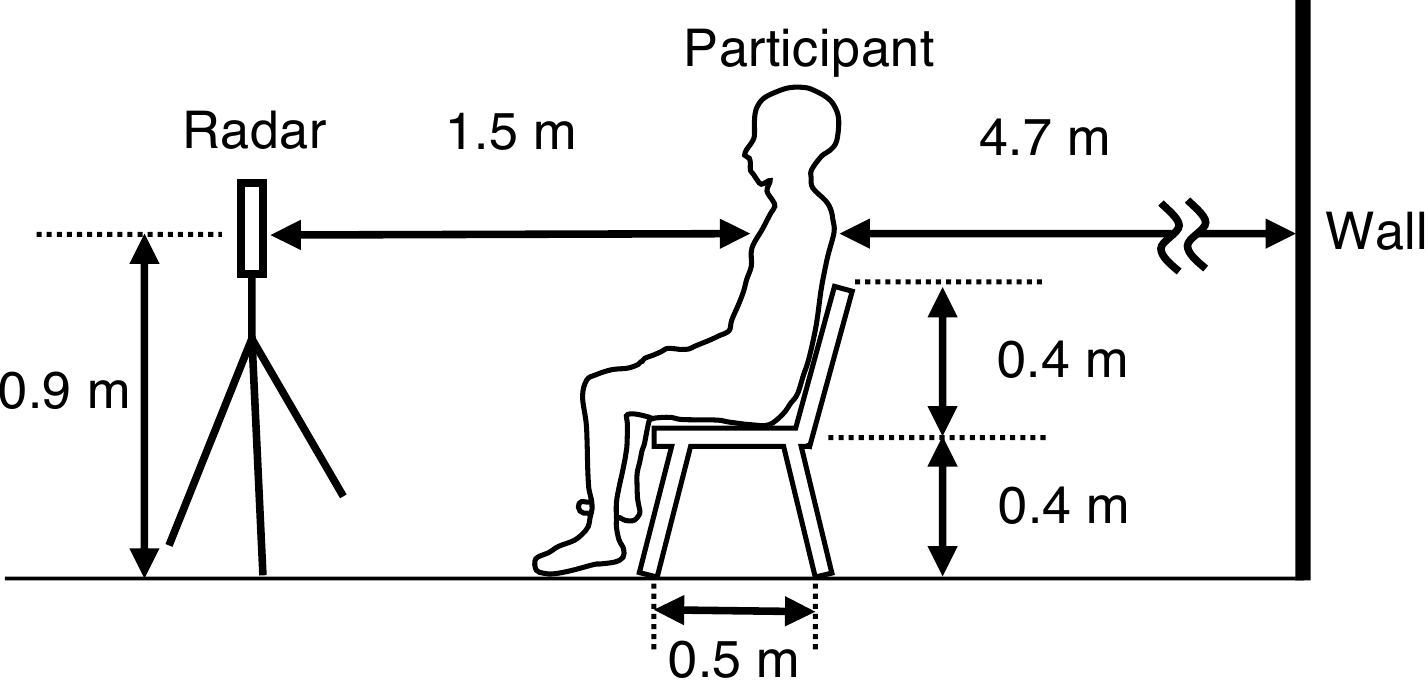}
            \subcaption{} \label{fig:schematic}
        \end{minipage}
        \caption{(a) Photograph and (b) schematic of the experimental setup showing the radar system and a seated participant.}
        \label{fig:scene_schematic}
    \end{figure}

In this experiment, measurements were conducted over five
    days with six healthy adult participants. Each day, measurements were taken five
    times per participant in the morning and another five times per participant in
    the afternoon, with each session lasting for $60.0~\mathrm{s}$. Therefore,
    the total number of measurements performed was $(5+5)\times 5 = 50$ times
    per each person, giving a total of 300 measurements.
    In other words, a total of 300 samples were taken. All participants were in good health with
    no history of respiratory problems or heart disease. The
    physical characteristics of each of the participants are listed in Table~\ref{tab:participants}.
    \begin{table}[tb!]
        \centering
        \caption{PHYSICAL CHARACTERISTICS OF EACH PARTICIPANT}
        \begin{tabular}{cccc}
            \toprule \textbf{Participant} & \textbf{Age (yr)} & \textbf{Height (m)} & \textbf{Weight (kg)} \\
            \midrule 1                     & 26           & 1.67                & 58.9                 \\
            2                              & 22           & 1.76                & 69.7                 \\
            3                              & 23           & 1.75                & 75.2                 \\
            4                              & 23           & 1.80                & 62.0                 \\
            5                              & 23           & 1.70                & 68.0                 \\
            6                              & 23           & 1.69                & 54.0                 \\
            \bottomrule
        \end{tabular}
        \label{tab:participants}
    \end{table}
    All measurements were conducted with each participant seated in front of the
    radar system at a distance of $1.5~\mathrm{m}$. The experimental setup is
    pictured in Fig.~\ref{fig:scene}, and a schematic that indicates the physical dimensions is shown
    in Fig.~\ref{fig:schematic}.

    The radar specifications are listed in Table~\ref{tab:1}. We used a
    frequency-modulated continuous-wave (FMCW) radar system with a center
    frequency of $79.0~\mathrm{GHz}$ and a bandwidth of $3.6~\mathrm{GHz}$. The antenna
    array consists of three transmitting and four receiving elements, which form
    a MIMO array that can be approximated virtually as a 12-element linear array.
    The range and the direction of arrival of the echo from the target are identified
    by performing a Fourier transform in terms of the fast time ~\cite{FMCW} and
    the virtual element number corresponding to the beamformer method~\cite{beamformer}.

    \begin{table}[t]
        \begin{center}
            \caption{SPECIFICATIONS OF THE RADAR SYSTEM}
            \label{tab:1}
            \begin{tabular}{cc}
                \toprule \textbf{Parameter}             & \textbf{Value}                \\
                \midrule Modulation                     & FMCW                          \\
                Center frequency                        & $79.0~\mathrm{GHz}$           \\
                Center wavelength                       & $3.8~\mathrm{mm}$             \\
                Bandwidth                               & $3.6~\mathrm{GHz}$            \\
                No. of transmitting (Tx) antennas       & $3$                           \\
                No. of receiving (Rx) antennas          & $4$                           \\
                Tx element spacing                      & $7.6~\mathrm{mm}$             \\
                Rx element spacing                      & $1.9~\mathrm{mm}$             \\
                Beamwidths of Tx elements (E-/H-planes) & $\pm 4^{\circ}/\pm33^{\circ}$ \\
                Beamwidths of Rx elements (E-/H-planes) & $\pm 4^{\circ}/\pm45^{\circ}$ \\
                Range resolution                        & $44.7~\mathrm{mm}$            \\
                Sampling frequency (slow-time)          & $100~\mathrm{Hz}$             \\
                \bottomrule
            \end{tabular}
        \end{center}
    \end{table}

\begin{figure*}
        \centering
        \begin{minipage}{0.32\linewidth}
            \centering
            \includegraphics[width=\linewidth]{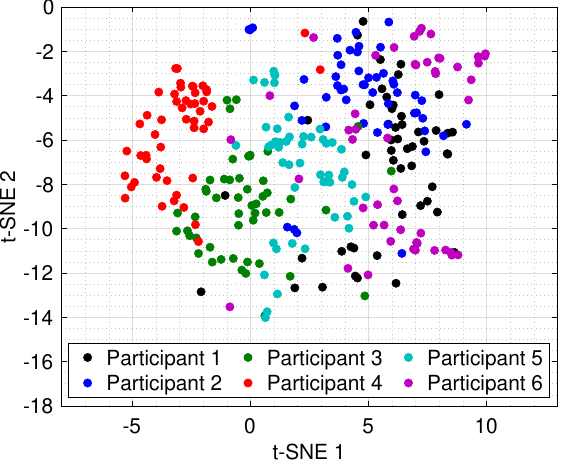}
            \subcaption{}
        \end{minipage}
        \begin{minipage}{0.32\linewidth}
            \centering
            \includegraphics[width=\linewidth]{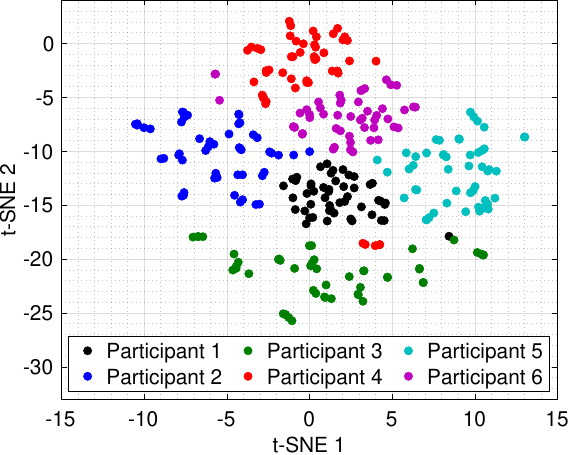}
            \subcaption{} \label{fig:tsne_vecB}
        \end{minipage}
        \begin{minipage}{0.32\linewidth}
            \centering
            \includegraphics[width=\linewidth]{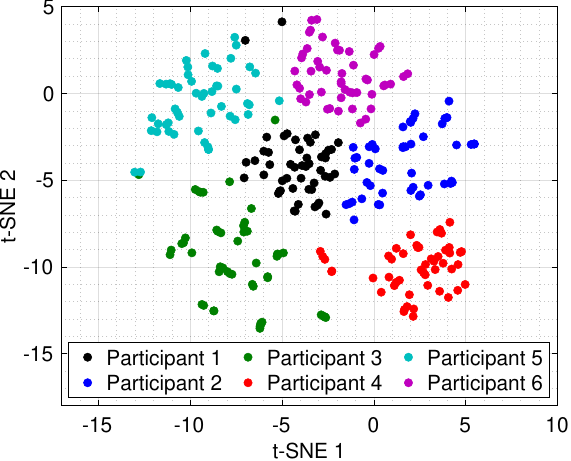}
            \subcaption{}
        \end{minipage}
        \caption{Two-dimensional t-SNE visualizations of the feature vectors (a) $\bm{r}_{\mathrm{resp}}$, (b) $\bm{r}_{\mathrm{hb}}$, and (c) $\bm{r}_{\mathrm{prop}}$ for the 60~s data samples acquired in experiment 1.}
        \label{fig:tsne}
    \end{figure*}

    The feature vectors $\bm{r}_{\mathrm{resp}}$, $\bm{r}_{\mathrm{hb}}$, and $\bm
    {r}_{\mathrm{prop}}$ are visualized by compressing them into two dimensions using
    the t-distributed stochastic neighbor embedding (t-SNE) method,~\cite{tSNE}
    as shown in Fig.~\ref{fig:tsne}, which shows that the feature vectors form
    clusters corresponding to the six participants. In particular, when compared
    with the other feature vectors, the clusters are separated clearly when using
    $\bm{r}_{\mathrm{prop}}$. In the $\bm{r}_{\mathrm{resp}}$ case, the cluster for
    participant~6 is erroneously split into two separate clusters and also overlaps
    with other clusters.

    The accuracy during individual identification was evaluated using a
    stratified $k$-fold cross-validation approach with $k=10$, thus ensuring that
    each subset maintained the class distribution of the original dataset; each
    cross-validation iteration used 270 samples for training and 30 samples for
    validation. As a single-fold validation, only data acquired from the same
    period can be used as validation data. For example, in the single-fold
    method, data measured on the morning of the first day are used as the validation
    data and the remaining data serve as training data.

    The three feature vectors $\bm{r}_{\mathrm{resp}}$, $\bm{r}_{\mathrm{hb}}$, and
    $\bm{r}_{\mathrm{prop}}$ are input into the three classifiers to compare their
    identification accuracies. Here, because both SVM and {$k$-NN} are binary classifiers,
    the one-versus-rest method is used to perform multi-class classification~\cite{o-vs-r}.
    In addition, the activation function used for the MLP output layer is the
    softmax function. The parameters for each classifier are optimized via a grid
    search. The optimization processes are performed for each of the three
    feature vectors $\bm{r}_{\mathrm{resp}}$, $\bm{r}_{\mathrm{hb}}$, and
    $\bm{r}_{\mathrm{prop}}$ individually. The parameters to be optimized and the
    values selected are given in Table~\ref{tab:parameters_ranges}.

    \begin{table*}
        [t]
        \centering
        \begin{threeparttable}
            \caption{OPTIMIZED PARAMETERS FOR EACH METHOD}
            \label{tab:parameters_ranges}
            \begin{tabular}{cccccc}
                \toprule \hspace*{3mm}\multirow{2}{*}{\textbf{Classifier}}\hspace*{3mm} & \hspace*{3mm}\multirow{2}{*}{\textbf{Parameters}}\hspace*{3mm} & \multirow{2}{*}{\textbf{Options}}\hspace*{3mm} & \multicolumn{3}{c}{\textbf{Feature vector}}\hspace*{3mm} \\
                \cmidrule(lr){4-6}                                                      &                                                                &                                                & A) $\bm{r}_{\mathrm{resp}}$                             & B) $\bm{r}_{\mathrm{hb}}$ & C) $\bm{r}_{\mathrm{prop}}$ \\
                \midrule 1) SVM                                                         & Kernel function                                                & Linear / Gaussian                              & Gaussian                                                & Gaussian                  & Gaussian                    \\
                \cmidrule(lr){1-6} \multirow{2}{*}{2) $k$-NN}                           & No. of neighbors                                               & $[1, 300]$                                     & 9                                                       & 28                        & 28                          \\
                                                                                        & Distance                                                       & Euclidean, Cityblock, Cosine                   & Cityblock                                               & Cosine                    & Cosine                      \\
                \cmidrule(lr){1-6} \multirow{4}{*}{3) MLP}                              & No. of layers                                                  & $[1, 3]$                                       & 2                                                       & 2                         & 2                           \\
                                                                                        & Size of \nth{1} layer                                          & $[1, 300]$                                     & 39                                                      & 47                        & 15                          \\
                                                                                        & Size of \nth{2} layer                                          & $[1, 300]$                                     & 19                                                      & 49                        & 15                          \\
                                                                                        & Activation function                                            & ReLU\tnote{\dag}, Sigmoid                      & ReLU\tnote{\dag}                                        & ReLU\tnote{\dag}          & ReLU\tnote{\dag}            \\
                \bottomrule
            \end{tabular}
            \begin{tablenotes}
                \item[\dag]: Rectified linear unit
            \end{tablenotes}
        \end{threeparttable}
    \end{table*}

\begin{figure*}[tb!]
        \begin{minipage}{0.99\linewidth}
            \centering
            \includegraphics[width=0.94\linewidth]{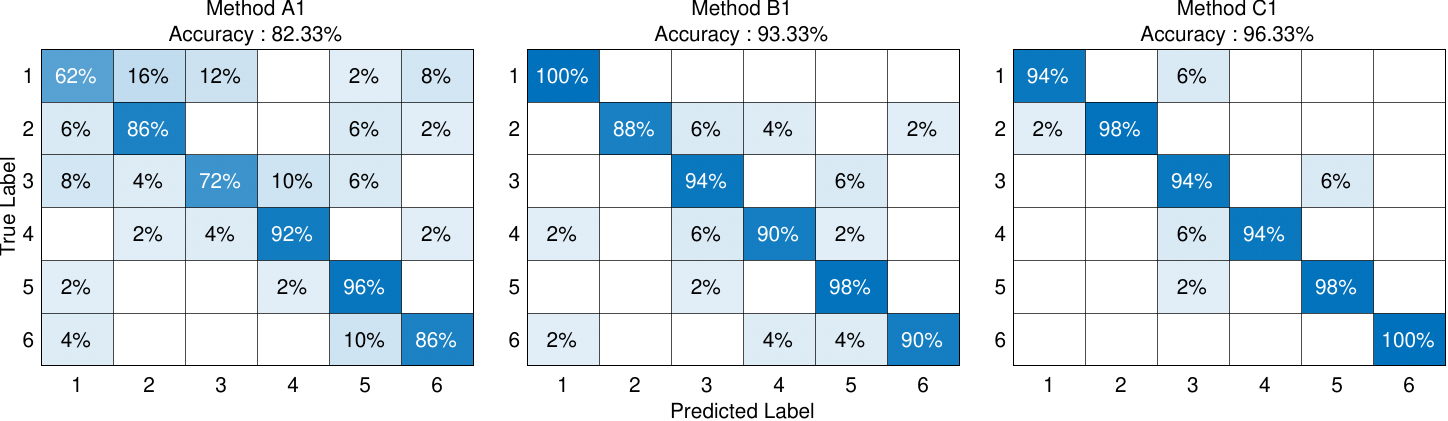}
            \subcaption{} \label{fig:confmat_svm}
        \end{minipage}
        \begin{minipage}{0.99\linewidth}
            \centering
            \vspace*{3mm}
            \includegraphics[width=0.94\linewidth]{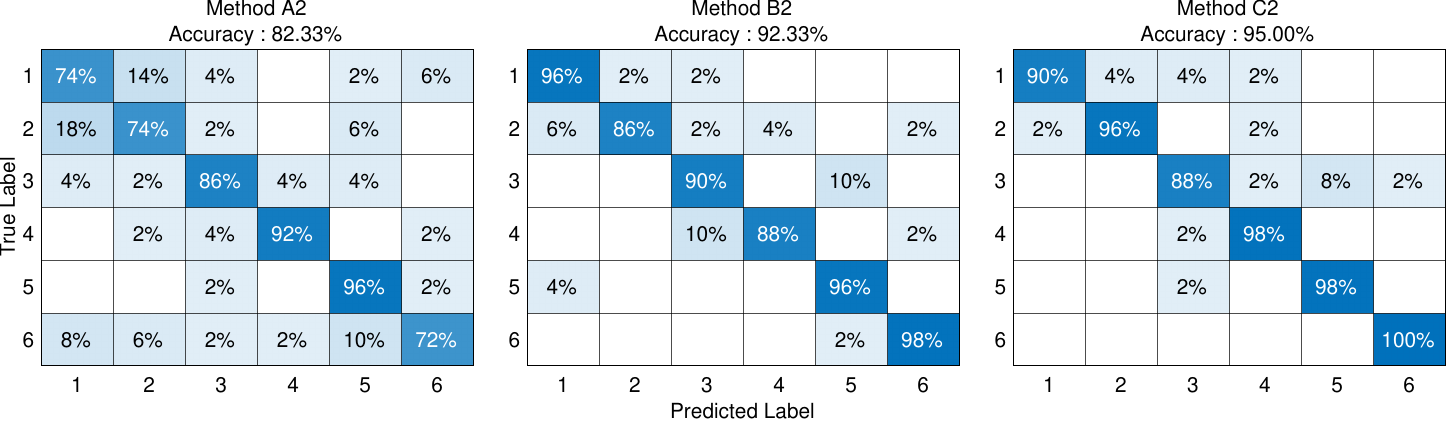}
            \subcaption{} \label{fig:confmat_knn}
        \end{minipage}
        \begin{minipage}{0.99\linewidth}
            \centering
            \vspace*{3mm}
            \includegraphics[width=0.94\linewidth]{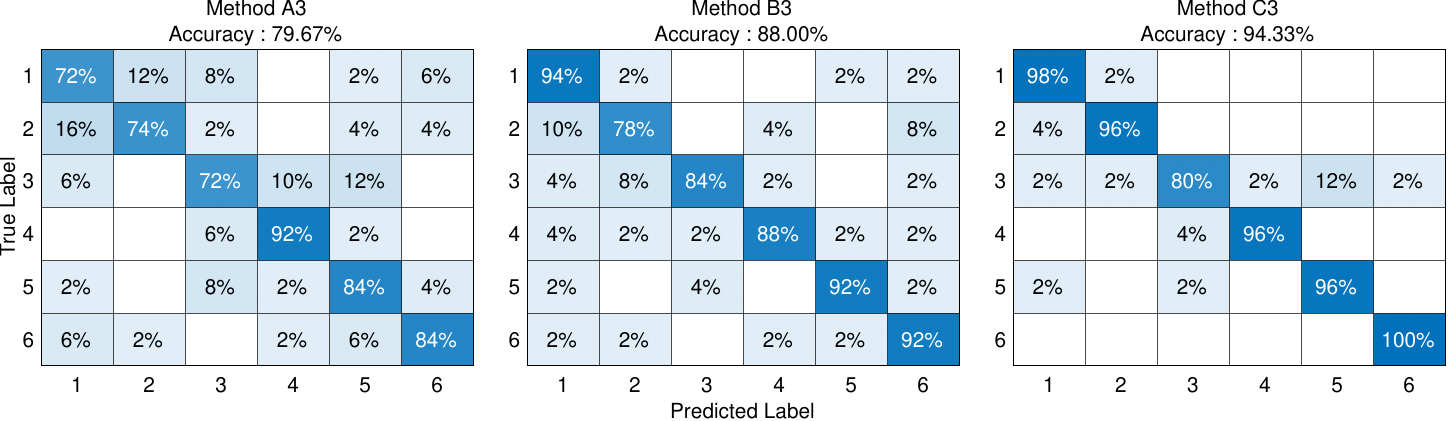}
            \subcaption{} \label{fig:confmat_mlp}
        \end{minipage}
        \centering
        \caption{Confusion matrices for each method (a) SVM, (b) $k$-NN, and (c) MLP for the 60~s data samples acquired in experiment 1.}
        \label{fig:confmat}
    \end{figure*}

    The confusion matrices obtained for each classifier are shown in Fig.~\ref{fig:confmat}.
    The accuracy indicates the proportion of correctly identified cases with
    respect to the total number of cases. When using $\bm{r}_{\mathrm{resp}}$,
    $\bm{r}_{\mathrm{hb}}$, and $\bm{r}_{\mathrm{prop}}$, these three
    classifiers achieved accuracies of approximately 80\%, 90\%, and 95\%,
    respectively. These results suggest that the heartbeat feature is superior to
    the respiratory feature when used in individual identification, and the
    combination of the respiratory and heartbeat components is superior to both
    the respiratory feature and the heartbeat feature when they are used
    separately. In Fig.~\ref{fig:confmat}, we see that method C1 realizes the highest
    accuracy of 96.33\% from among the nine methods.

\begin{figure*}[tb!]
        \centering
        \begin{minipage}{0.3\linewidth}
            \centering
            \includegraphics[width=\linewidth]{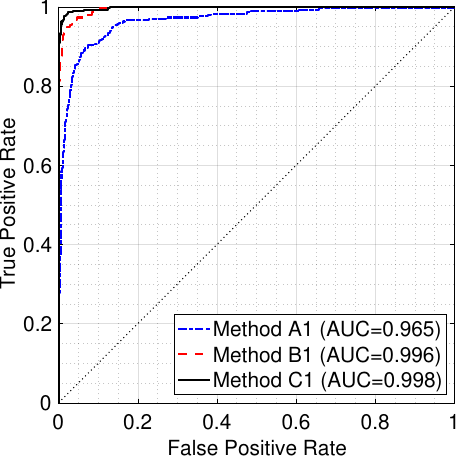}
            \subcaption{}
        \end{minipage}
        \hspace*{3mm}
        \begin{minipage}{0.3\linewidth}
            \centering
            \includegraphics[width=\linewidth]{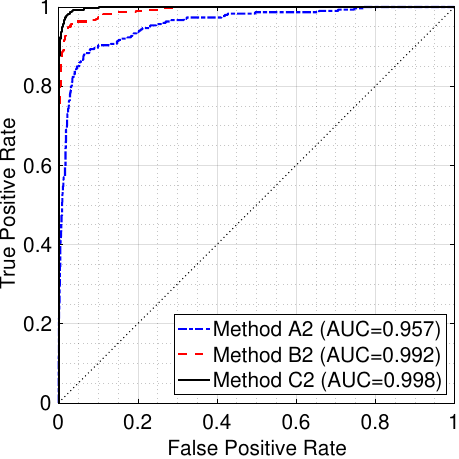}
            \subcaption{}
        \end{minipage}
        \hspace*{3mm}
        \begin{minipage}{0.3\linewidth}
            \centering
            \includegraphics[width=\linewidth]{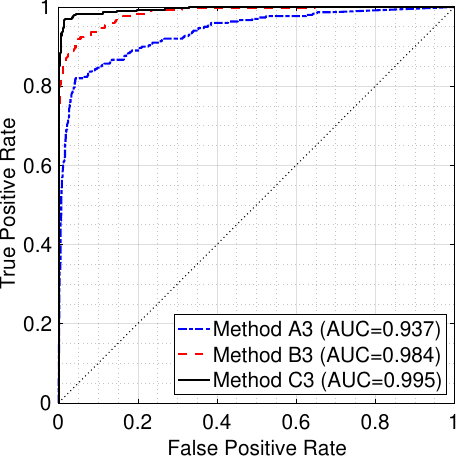}
            \subcaption{}
        \end{minipage}
        \caption{ROC curves for (a) SVM, (b) $k$-NN, and (c) MLP for the 60~s data samples acquired in experiment 1.}
        \label{fig:roc}
    \end{figure*}

    Fig.~\ref{fig:roc} presents the receiver operating characteristic (ROC)
    curves and the corresponding area under the curve (AUC) results for each
    classifier. Note that the ROC curves shown in Fig.~\ref{fig:roc} are the
    macro-averages of the one-versus-rest ROC curves for each class. Fig.~\ref{fig:roc}
    shows that when $\bm{r}_{\mathrm{prop}}$ is used, the AUC for all classifiers
    reaches 0.99 and is higher than that realized when using $\bm{r}_{\mathrm{resp}}$
    and $\bm{r}_{\mathrm{hb}}$ separately.

    Table~\ref{tab:3} presents the F$_{1}$ scores and the AUCs for each
    participant when $\bm{r}_{\mathrm{prop}}$ is used as the input, along with their
    respective macro-averages. From Table \ref{tab:3}, we see that the SVM achieves
    the highest F$_{1}$ scores and AUCs among the classifiers. The AUC exceeds a
    macro average of 0.99 for all classifiers, thus indicating that each
    classifier achieves comparable accuracy to the others. This indicates that the
    proposed features enable high-accuracy identification, regardless of the
    classifier that is used.

    \begin{table}[tb!]
        \centering
        \caption{F$_{1}$ SCORE AND AUC FOR EACH PARTICIPANT USING $\bm{r}_{\mathrm{prop}}$
        IN EXP.~1}
        \label{tab:3}
        \begin{tabular}{cccc}
            \toprule \hspace*{2mm}\textbf{Classifier}\hspace*{2mm} & \hspace*{2mm}\textbf{Participant}\hspace*{2mm} & \hspace*{2mm}\textbf{F$_{1}$ score}\hspace*{2mm} & \hspace*{2mm}\textbf{AUC}\hspace*{2mm} \\
            \midrule \multirow{7}{*}{1) SVM (method C1)}           & 1                                              & 0.959                                            & 0.994                                  \\
                                                                   & 2                                              & 0.990                                            & 0.999                                  \\
                                                                   & 3                                              & 0.904                                            & 0.995                                  \\
                                                                   & 4                                              & 0.969                                            & 1.000                                  \\
                                                                   & 5                                              & 0.961                                            & 0.999                                  \\
                                                                   & 6                                              & 1.000                                            & 1.000                                  \\
            \cmidrule(lr){2-4}                                     & macro-ave.                                     & 0.964                                            & 0.998                                  \\
            \midrule \multirow{7}{*}{2) $k$-NN (method C2)}        & 1                                              & 0.938                                            & 0.998                                  \\
                                                                   & 2                                              & 0.960                                            & 0.997                                  \\
                                                                   & 3                                              & 0.898                                            & 0.996                                  \\
                                                                   & 4                                              & 0.961                                            & 0.998                                  \\
                                                                   & 5                                              & 0.952                                            & 0.999                                  \\
                                                                   & 6                                              & 0.990                                            & 1.000                                  \\
            \cmidrule(lr){2-4}                                     & macro-ave.                                     & 0.950                                            & 0.998                                  \\
            \midrule \multirow{7}{*}{3) MLP (method C3)}           & 1                                              & 0.951                                            & 0.999                                  \\
                                                                   & 2                                              & 0.960                                            & 0.997                                  \\
                                                                   & 3                                              & 0.860                                            & 0.979                                  \\
                                                                   & 4                                              & 0.970                                            & 0.999                                  \\
                                                                   & 5                                              & 0.923                                            & 0.997                                  \\
                                                                   & 6                                              & 0.990                                            & 0.999                                  \\
            \cmidrule(lr){2-4}                                     & macro-ave.                                     & 0.942                                            & 0.995                                  \\
            \bottomrule
        \end{tabular}
        \label{tab:performance}
    \end{table}

    Next, to investigate the individual identification performance for short data samples, we divide each 60~s data sample into 12 short data samples with a length of 5.0~s, thus generating 3,600 samples in total. For these shorter samples, we evaluate the performance for the heartbeat feature $\bm{r}_{\mathrm{hb}}$ only (corresponding to methods B1, B2, and B3) because the data sample length of 5.0~s is not sufficient to capture respiratory features, which have a typical interval of between 3~s and 5~s.

\begin{figure}
        \centering
        \includegraphics[width=0.7\linewidth]{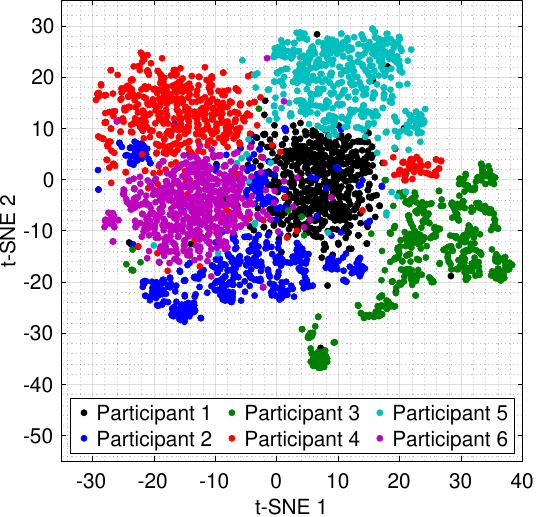}
        \caption{Two-dimensional t-SNE visualizations of the feature vector $\bm{r}_{\mathrm{hb}}$ for the 5~s data samples acquired in experiment 1.}
        \label{fig:exp1_tsne_5s}
    \end{figure}

\begin{figure*}
        \centering
        \begin{minipage}{0.32\linewidth}
            \centering
            \includegraphics[width=0.9\linewidth]{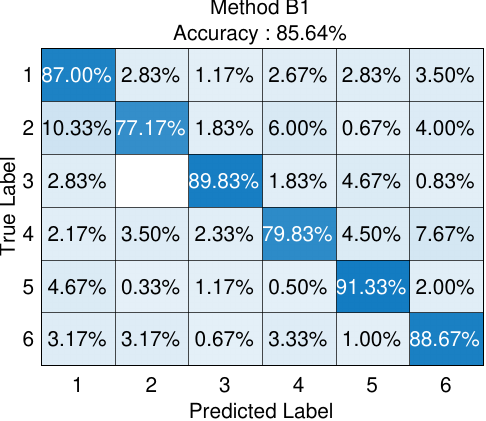}
            \subcaption{}
        \end{minipage}
        \begin{minipage}{0.32\linewidth}
            \centering
            \includegraphics[width=0.9\linewidth]{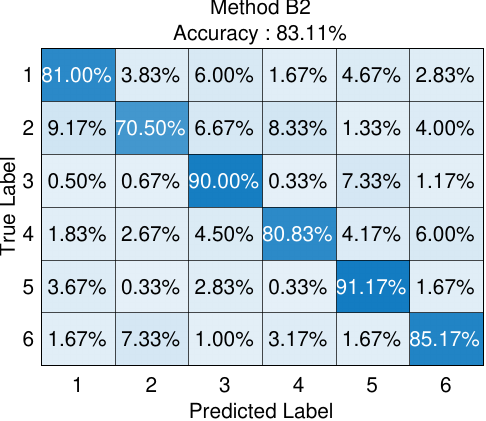}
            \subcaption{}
        \end{minipage}
        \begin{minipage}{0.32\linewidth}
            \centering
            \includegraphics[width=0.9\linewidth]{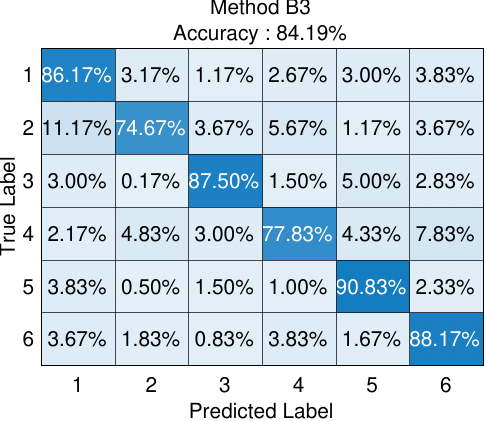}
            \subcaption{}
        \end{minipage}
        \caption{Confusion matrices for each method: (a) SVM, (b) $k$-NN, and (c) MLP for the 5~s data samples acquired in experiment 1.}
        \label{fig:confmat_5s}
    \end{figure*}

    The feature vector $\bm{r}_{\mathrm{hb}}$ for the short data samples is visualized in two-dimensional space using the t-SNE method in Fig.~\ref{fig:exp1_tsne_5s}. Although the figure shows six clusters corresponding to the six participants, these clusters overlap severely when compared with those shown in Fig.~\ref{fig:tsne_vecB}; this is an indication of the performance degradation caused by the limited data sample length.
    The confusion matrices obtained using the stratified $k$-fold cross-validation approach are shown in Fig.~\ref{fig:confmat_5s} for $k=10$ for the 5~s data samples. The accuracies obtained for methods B1, B2, and B3 with the 5~s data samples were 85.64\%, 83.11\%, and 84.19\%, respectively, showing that the performance has been reduced by 7.69, 9.22, and 3.81 points when compared with the accuracies of 93.33\%, 92.33\%, and 88.00\% for B1, B2, and B3, respectively, obtained with the 60~s data samples. We also confirmed that the AUCs obtained for methods B1, B2, and B3 with the 5~s data samples were 0.979, 0.968, and 0.962, respectively, which were lower than the AUCs obtained with the 60~s data samples by 0.017, 0.024, and 0.022, respectively.

    \subsection{Experiment 2: Public Dataset of Thirty Participants}
    Next, we evaluated the accuracy of the proposed method using a public dataset
    provided by Schellenberger {\em et al.}~\cite{https://doi.org/10.1038/s41597-020-00629-5},
    where they used a continuous-wave (CW) radar system with an operating
    frequency of $24~\mathrm{GHz}$ to collect echoes from a total of 30
    individuals, comprising 14 male and 16 female participants. The echoes were
    recorded with a sampling frequency of $2.0~\mathrm{kHz}$, which was then downsampled
    to $100~\mathrm{Hz}$ to make the conditions consistent with those of experiment
    1. The measurement time varied for each participant and ranged from approximately
    $600~\mathrm{s}$ to $700~\mathrm{s}$. In this study, we divided the data into
    segments with a length of $T_{0}$ and examined two cases, where
    $T_{0}=60.0~\mathrm{s}$ and $T_{0}=5.0~\mathrm{s}$; therefore,
    totals of 306 and 3,795 samples were taken, respectively. A feature vector
    was then generated from each sample.

\begin{figure*}[tb!]
        \centering
        \begin{minipage}{0.49\linewidth}
            \centering
            \includegraphics[width=0.60\linewidth]{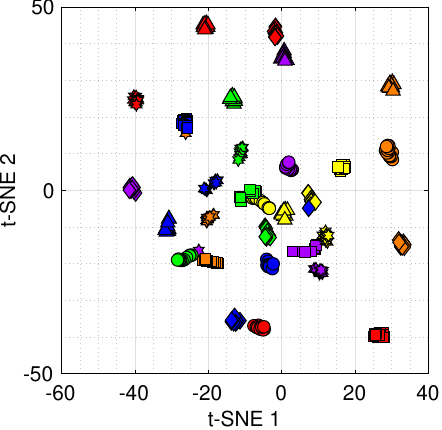}
            \subcaption{}
        \end{minipage}
        \begin{minipage}{0.49\linewidth}
            \centering
            \includegraphics[width=0.60\linewidth]{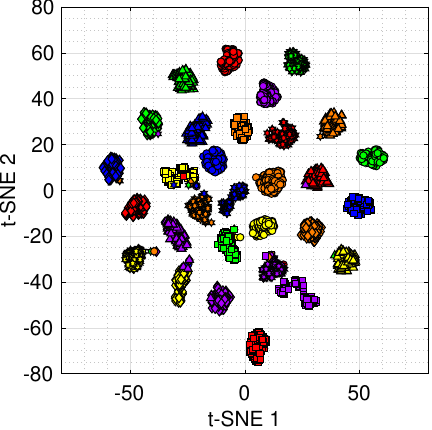}
            \subcaption{}
        \end{minipage}
        \caption{Two-dimensional plots of feature vectors $\bm{r}_{\mathrm{hb}}$
        compressed by t-SNE when using the public dataset with data lengths (a)
        $T_{0}=60.0$ s and (b) $T_{0}=5.0$ s in experiment 2.}
        \label{fig:dataset_tsne}
    \end{figure*}

\begin{figure*}[tb!]
        \centering
        \begin{minipage}{0.99\linewidth}
            \includegraphics[width=0.95\linewidth]{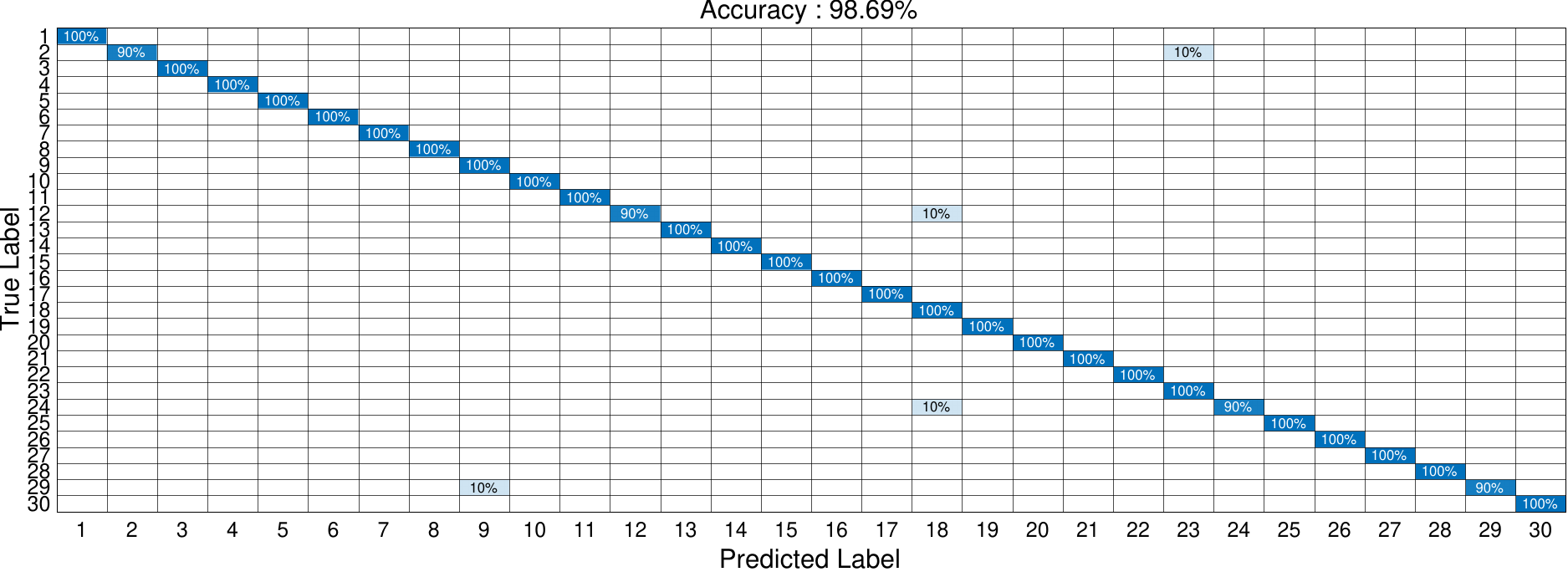}
            \subcaption{} \label{fig:60s_confmat}
        \end{minipage}
        \begin{minipage}{0.99\linewidth}
            \vspace*{3mm}
            \includegraphics[width=0.95\linewidth]{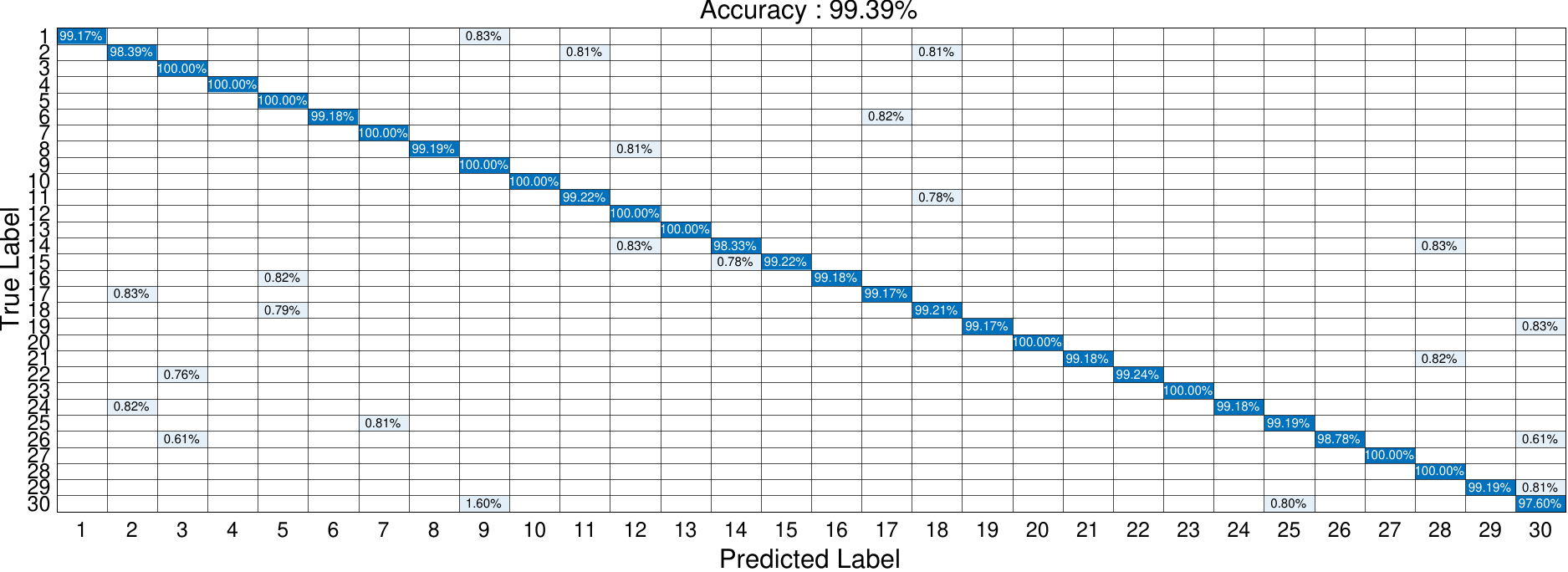}
            \subcaption{} \label{fig:5s_confmat}
        \end{minipage}
        \caption{Confusion matrices for experiment 2 with a data length (a)
        $T_{0}=60~\mathrm{s}$ and (b) $T_{0}=5~\mathrm{s}$.}
        \label{fig:confmat_public}
    \end{figure*}

    Fig.~\ref{fig:dataset_tsne} visualizes two-dimensional plots of the feature vector
    $\bm{r}_{\mathrm{hb}}$ as obtained from data segments with $T_{0}=60.0~\mathrm{s}$
    and $T_{0}=5.0~\mathrm{s}$ when compressed using t-SNE. In both cases, it was
    observed that almost all plots were clearly separated into distinct clusters
    corresponding to the 30 participants. Next, the feature vector $\bm{r}_{\mathrm{hb}}$
    obtained was input into the SVM and the identification accuracy was evaluated
    by performing a stratified $k$-fold cross-validation with $k=5$; the
    resulting confusion matrices are shown in Fig.~\ref{fig:confmat_public}.
    First, when $T_{0}=60.0~\mathrm{s}$, 100\% accuracy was achieved for all but
    four participants, and the overall accuracy was 98.69\%, thus indicating
    that individual identification was achieved successfully. Next, when
    $T_{0}=5.0~\mathrm{s}$, the overall accuracy realized was 99.39\%, which also
    represents high-precision identification.

    In experiment 1, the accuracy of method B2 was only 93.33\%, whereas in
    experiment 2, the same method B2 achieved accuracy of 98.69\%, partly because
    the public dataset was collected within a day and was thus unaffected by daily
    variations in the physiological signals. We should also note that the
    accuracy obtained for $T_{0}=5.0~\mathrm{s}$ was almost the same as the
    accuracy for $T_{0}=60.0~\mathrm{s}$, which may indicate that long measurements
    are not necessary during individual identification if the daily variations in
    the physiological signals can be neglected.

    \section{Discussion}
    \subsection{Effect of Data Length}
    In experiment 1, the accuracy for the 5~s data samples was lower than that for the 60~s data samples.
    In contrast, in experiment 2, the accuracy remained high even when the 5~s data samples were used.
    The difference between the two experiments is that the measurements in experiment~1 were taken repeatedly, twice a day over five days, and thus the respiratory and heartbeat features were not completely consistent over the datasets that were acquired on different days.
    In contrast, the measurements in experiment 2 were all performed over a single day, and thus the respiratory and heartbeat features remained relatively consistent over the datasets that were used for training and testing.
    Note that in experiment 1, when datasets acquired in the morning on day 1 were used for testing, we then selected training datasets acquired in the afternoon on day 1 in addition to those from days 2, 3, 4, and 5; we thus avoided using datasets that had been acquired in the same half-day session for both training and testing, which made the evaluation conditions used in experiment 1 difficult but practical.

    In addition, the measurement setups used in experiments 1 and 2 were different; in experiment 1, the participants were seated, whereas in experiment 2, the participants were lying down on a bed.
    As a result, there were larger participant body movements in experiment 1 than in experiment 2, which made problem setting in experiment 1 more challenging.
    For these reasons, individual identifications in experiment 1 required longer data lengths than in experiment 2.
    In future work, it will be important to develop a method for individual identification when using short data samples, even when the participants are seated and the datasets for training and testing are acquired on different days.

    \subsection{Effect of Background Noise}
    Next, we investigate the effect of background noise on the identification performance by applying the methods to data with a low signal-to-noise power ratio (SNR).
    The low-SNR data required were generated by adding white Gaussian noise to the measured data from in experiment 1, where this noise was approximated using a random complex-valued sequence following a normal distribution.
    Note here that the SNR of the data acquired in experiment 1 was originally 36.9~dB.

    Table~\ref{tab:snr} lists the accuracy and the AUC for each method with SNRs of 10.0, 20.0, and 30.0~dB. Using the original data with the high SNR of 36.9~dB, the average accuracy obtained for methods A1, A2, and A3 was 81.67\%; when using the data with the low SNR of 10.0~dB, the average accuracy for methods A1, A2, and A3 was 80.56\%, which indicates that a similar level of accuracy is maintained when using the respiratory features, even with the low-SNR data.
    For example, the accuracies for method A1 were 81.67\%, 82.67\%, and 80.67\% for the SNRs of 10.0~dB, 20.0~dB, and 30.0~dB, respectively.

    In contrast, when using the original data with the SNR of 36.9~dB, the average accuracy for methods B1, B2, and B3 was 91.22\%; when using the data with the SNR of 10.0~dB, the average accuracy for methods B1, B2, and B3 was 78.33\%, indicating that the averaged accuracy decreased by 12.89 points.
    For example, the accuracies for method B1 when using the data with the SNRs of 10.0~dB, 20.0~dB, and 30.0~dB were 79.33\%, 85.00\%, and 87.33\%, respectively; these values were lower than the accuracies obtained for the data with the high SNR of 36.9~dB by 14.00, 8.33, and 6.00 points, respectively.
    These results indicate that the methods that use the heartbeat features (i.e., methods B1, B2, and B3) are affected more severely by the background noise when compared with the the methods that use respiratory features (i.e., methods A1, A2, and A3).
    This is attributed to the fact that the body displacement caused by the heartbeat is much smaller than that caused by respiration; thus, the heartbeat features are more susceptible to noise.

    \begin{table}[tb]
        \centering
        \caption{ACCURACY AND AUC VALUES FOR LOW SNR CONDITIONS}
        \begin{tabular}{lcccccc}
            \toprule \multirow{2}{*}{\bf Method}  & \multicolumn{3}{c}{\bf Accuracy (\%)} & \multicolumn{3}{c}{\bf AUC} \\
            \cmidrule(lr){2-4} \cmidrule(lr){5-7} &10~dB& 20~dB& 30~dB                                 & 10~dB                      & 20~dB & 30~dB \\
            \midrule A1                    & 81.67 & 82.67                                 & 80.67 & 0.962                    & 0.965 & 0.968 \\
            A2                             & 78.67 & 79.00                                 & 78.67 & 0.948                    & 0.958 & 0.944 \\
            A3                             & 81.33 & 78.67                                 & 80.33 & 0.935                    & 0.930 & 0.947 \\
            \cmidrule(lr){1-7} B1          & 79.33 & 85.00                                 & 87.33 & 0.951                    & 0.972 & 0.981 \\
            B2                             & 75.33 & 81.67                                 & 86.33 & 0.931                    & 0.968 & 0.978 \\
            B3                             & 80.33 & 83.33                                 & 83.67 & 0.938                    & 0.963 & 0.965 \\
            \cmidrule(lr){1-7} C1          & 88.67 & 92.00                                 & 92.33 & 0.990                    & 0.992 & 0.993 \\
            C2                             & 86.33 & 90.33                                 & 93.00 & 0.975                    & 0.988 & 0.993 \\
            C3                             & 87.67 & 91.33                                 & 92.33 & 0.978                    & 0.985 & 0.992 \\
            \bottomrule
        \end{tabular}
        \label{tab:snr}
    \end{table}

\subsection{Effects of Clothing and Temperature}
    When the measurements were conducted in experiments~1 and 2, all participants were dressed.
    It has been reported previously that the reflection from clothing is small enough to be considered negligible in measurements performed using millimeter-wave radar~\cite{clothes1, clothes2, clothes3}. Therefore, it is unnecessary to consider the effects of vibration of clothes in processing the radar data. Even if there is an effect from the participant's clothes, our proposed method achieved individual identification successfully for the dressed participants, which indicates the effectiveness of the radar-based approach.

    The radar-based physiological measurement principle is based on the detection of small displacements caused by the subject's heartbeat, respiration, and body motion.
    Therefore, the radar measurements are affected by neither the sample temperature nor the room temperature.
    However, as the participant's body temperature increases, their respiratory rate and heart rate also increase in general \cite{Kirschen2020}, which may then affect the method's accuracy when identifying multiple individuals.

    \subsection{Performance Under Different Conditions}
    In this section, we evaluate the individual identification performance under eight different conditions, designated $\alpha$1, $\alpha$2, $\beta$1, $\beta$2, $\gamma$1, $\gamma$2, $\delta$, and $\varepsilon$, which are defined in Table~\ref{tab:condition}.
    We performed additional radar measurements five times for both participants~3 and 4 from experiment~1, and the additional data were used for testing only, and not for training; the classifiers were trained using the original data from experiment~1.

    Fig.~\ref{fig:tsne_add} shows the feature vectors $\bm{r}_{\mathrm{resp}}$, $\bm{r}_{\mathrm{hb}}$, and $\bm{r}_{\mathrm{prop}}$ that were obtained from the additional measurements and then compressed into two dimensions using the t-SNE method; Table~\ref{tab:coff1} presents the confusion matrices for each of the eight conditions.
    The additional radar measurements were performed nine months after the original radar measurements described in Section~\ref{subsec:exp1}.

    \subsubsection{Condition $\alpha$: Target Distance}
    First, we evaluated the performance at various distances between the radar and the target person.
    Conditions $\alpha$1 and $\alpha$2 correspond to distances of 1.0~m and 3.0~m, respectively, although the distance was originally set at 1.5~m in experiment~1.

    Table~\ref{tab:coff1} shows that although the accuracy under condition $\alpha$1 was marginally low, the accuracy under condition $\alpha$2 was high, and Fig.~\ref{fig:tsne_add} shows that the feature vectors $\bm{r}_{\mathrm{resp}}$ are located far away from the corresponding cluster under condition $\alpha$1.
    The performance of method A1 in particular is affected negatively under condition $\alpha$1.
    These results may be caused by the shift in the radar reflection points on the target person's body, depending on the distance to the target, which leads to discrepancies in the respiratory feature vectors that are used for training and testing.

    \subsubsection{Condition $\beta$: Radar Height}
    Next, we evaluated the influence of the radar height from the floor; under conditions $\beta$1 and $\beta$2, the radar was installed at heights of 0.8~m and 1.0~m, respectively, whereas the radar was originally installed at a height of 0.9~m from the floor.
    As Table~\ref{tab:coff1} shows, the accuracy of method A1 is low under condition $\beta$1, whereas the accuracy of method B1 remains high under both conditions $\beta$1 and $\beta$2.
    In addition, as illustrated in Fig.~\ref{fig:tsne_add}, the respiratory feature vector $\bm{r}_\mathrm{resp}$ under condition $\beta$1 is located near the original cluster, which indicates that the individual characteristics of each participant were captured under both conditions. Therefore, it can be deduced that the radar height does not affect the individual identification performance significantly.

    \subsubsection{Condition $\gamma$: Seating Direction}
\begin{figure}
        \centering
        \begin{minipage}{0.49\linewidth}
            \includegraphics[width=\linewidth]{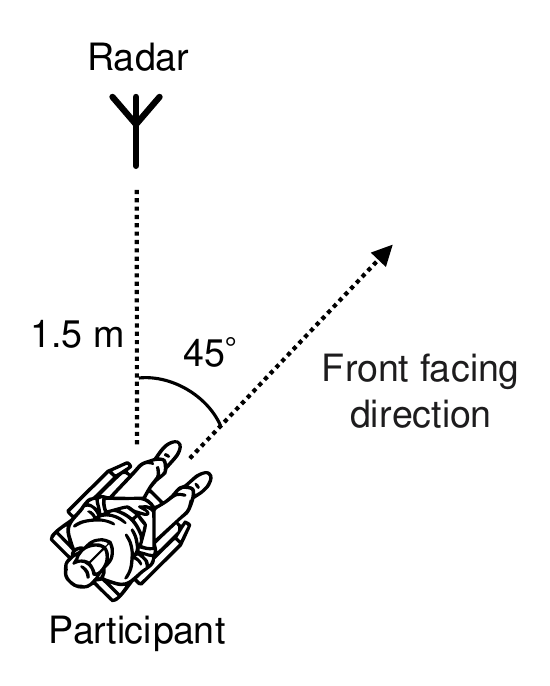}
            \subcaption{}
        \end{minipage}
        \begin{minipage}{0.49\linewidth}
            \includegraphics[width=\linewidth]{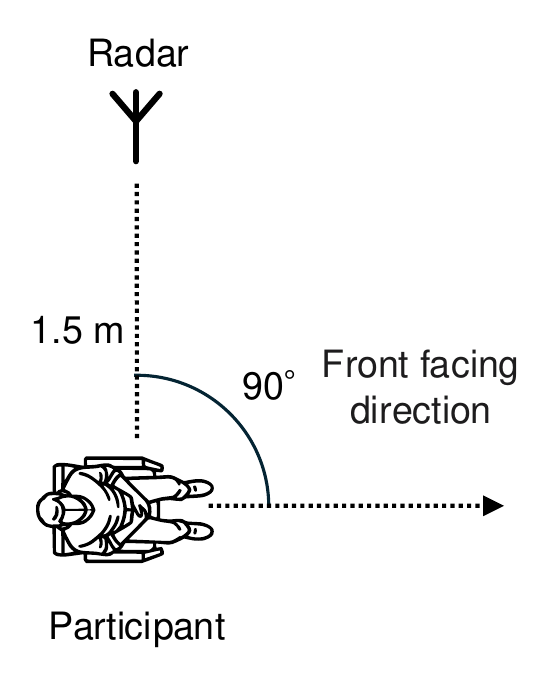}
            \subcaption{}
        \end{minipage}
        \caption{Schematics of the experimental setup with the radar system and seated participants under (a) condition $\bm{\gamma}$1 and (b) condition $\bm{\gamma}$2.}
        \label{fig:angle_fig}
    \end{figure}

    Next, we evaluated the effect of the target person's seating direction on the identification performance.
    Under conditions $\gamma$1 and $\gamma$2, we performed measurements for participants when seated facing at angles of $45^\circ$ and $90^\circ$, respectively, relative to the radar system; the corresponding experimental setups are illustrated in Fig.~\ref{fig:angle_fig}.
    The results in Table~\ref{tab:coff1} show that the accuracy decreases under both conditions $\gamma$1 and $\gamma$2, with none of the methods achieving high performances.
    Fig.~\ref{fig:tsne_add} also indicates that the feature vectors determined under conditions $\gamma$1 and $\gamma$2 do not capture the individual features of each participant.

    The results above can be explained as follows.
    The body displacements caused by the respiratory and heartbeat motions are particularly large at the front of the torso, whereas the displacements measured under conditions $\gamma$1 and $\gamma$2 are smaller because the echo is likely to come from a different side of the torso or from the shoulder/arm area.
    The performance evaluation results under conditions $\gamma$1 and $\gamma$2 indicate that the seating direction can affect the individual identification accuracy significantly.

    \subsubsection{Condition $\delta$: Wall Interference}
    In condition $\delta$, we evaluated the effects of clutter reflected from the mortar wall behind the participant.
    As shown in Fig.~\ref{fig:scene_schematic}, the wall was located comparatively far behind the target person in the original measurement setup used in experiment 1.
    We therefore performed radar measurements where each participant's back is in contact with the wall, corresponding to a distance of 0~m from the wall.
    Under this condition, although the accuracy decreases in the manner shown in Table~\ref{tab:coff1}, Fig.~\ref{fig:tsne_add_vecC} demonstrates that the feature vector $\bm{r}_\mathrm{prop}$ remains close to the original cluster because the radar system used in experiment~1 has a wide bandwidth of 3.6~GHz, which provides a range resolution of 41.7~mm that is sufficient to prevent the wall clutter from contaminating the target person's echo.

    \subsubsection{Condition $\varepsilon$: Exhaustion and Sweating}
\begin{figure*}
        \centering
        \begin{minipage}{0.32\linewidth}
            \includegraphics[width=\linewidth]{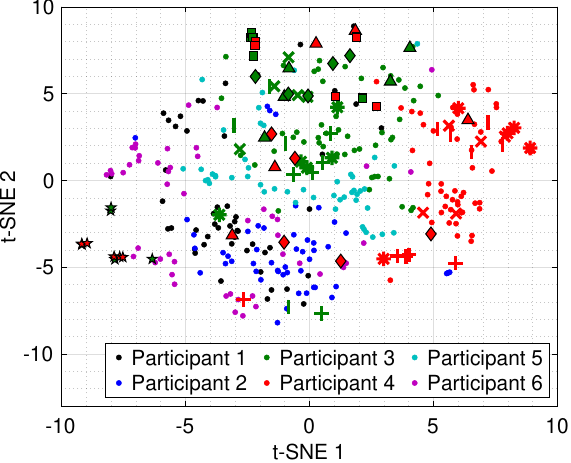}
            \subcaption{}
        \end{minipage}
        \begin{minipage}{0.32\linewidth}
            \includegraphics[width=\linewidth]{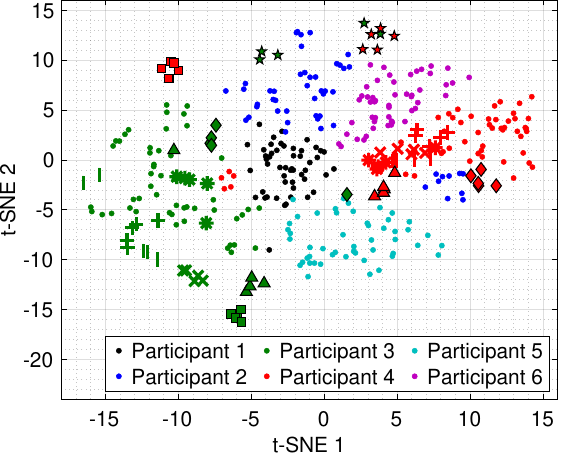}
            \subcaption{}
        \end{minipage}
        \begin{minipage}{0.32\linewidth}
            \includegraphics[width=\linewidth]{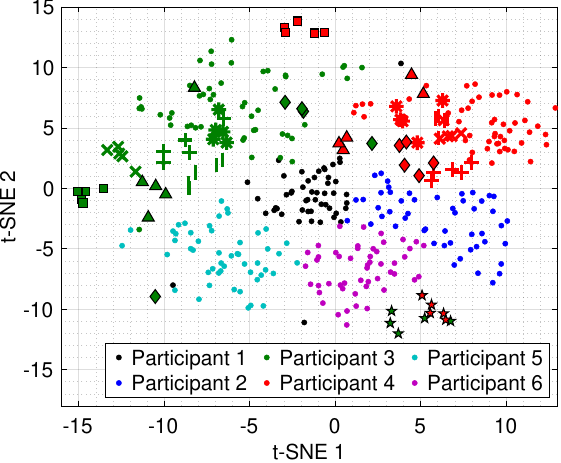}
            \subcaption{}
            \label{fig:tsne_add_vecC}
        \end{minipage}
        \caption{Two-dimensional t-SNE visualizations of the feature vectors (a) $\bm{r}_{\mathrm{resp}}$, (b) $\bm{r}_{\mathrm{hb}}$, and (c) $\bm{r}_{\mathrm{prop}}$ under each condition (Conditions: $\bm{\alpha}$1:~+; $\bm{\alpha}$2:~$\bm{\ast}$; $\bm{\beta}$1:~$|$; $\bm{\beta}$2:~x; $\bm{\gamma}$1:~$\triangle$; $\bm{\gamma}$2:~$\square$; $\bm{\delta}$:~$\Diamond$; $\bm{\varepsilon}$:~\ding{73}).}
        \label{fig:tsne_add}
    \end{figure*}

    In condition $\varepsilon$, we evaluated the performance when the participant is exhausted and sweating; immediately before the measurements were taken, the participants were instructed to run up 8 flights of stairs.
    During the measurements, the participants were out of breath and sweating as a result.
    Under condition $\varepsilon$, the accuracy was low for all methods, as shown in Table~\ref{tab:coff1}.
    Fig.~\ref{fig:tsne_add_vecC} shows that the feature vectors $\bm{r}_\mathrm{prop}$ are all far from the original clusters, which indicates that the unique individual features were not maintained after the intense exercise.
    The significant increases observed in both the respiratory rate and the heart rate after exercise are likely to have a considerable impact on the performance of the proposed method.
    It can be concluded from these results that the proposed method is most suitable for use when the target person is at rest.

    \begin{table}[bt!]
    \centering
    \caption{DEFINITIONS OF MEASUREMENT CONDITIONS}
    \begin{tabular}{lcc}
        \toprule \bf Condition          & \bf Parameter                    & \bf Value \\
        \midrule Condition $\alpha$1           & \multirow{2}{*}{Target distance} & 1.0~m     \\
        Condition $\alpha$2                    &                                  & 3.0~m     \\
        \cmidrule(rl){1-3} Condition $\beta$1 & \multirow{2}{*}{Radar height}    & 0.8~m     \\
        Condition $\beta$2                    &                                  & 1.0~m     \\
        \cmidrule(rl){1-3} Condition $\gamma$1 & \multirow{2}{*}{Seating direction} & $45^\circ$   \\
        Condition $\gamma$2                    &                                  & $90^\circ$   \\
        \cmidrule(rl){1-3} Condition $\delta$  & Wall interference                 & -         \\
        \cmidrule(rl){1-3} Condition $\varepsilon$  & Exhaustion and sweating     & -         \\
        \bottomrule
    \end{tabular}
    \label{tab:condition}
    \end{table}
    \begin{table*}[bt!]
        \centering
        \caption{CONFUSION MATRICES UNDER DIFFERENT MEASUREMENT CONDITIONS}
        \label{tab:coff1}
        \centering
        \begin{tabular}{@{}r@{}ccccccccccccccccccccccccccccc}
            \toprule                                                                                                                                                         &                                                  & \multicolumn{13}{c}{\bf Predicted Label} &                                                  & \multicolumn{13}{c}{\bf Predicted Label} \\
            \multicolumn{2}{c}{\multirow{3}{*}{\bf True Label}}                                                                                                              & \multicolumn{6}{c}{{\bf Condition $\bm{\alpha}$1} (1.0~m)}   &                                          & \multicolumn{6}{c}{{\bf Condition $\bm{\alpha}$2} (3.0~m)}   &                                         & \multicolumn{6}{c}{{\bf Condition $\bm{\beta}$1} (0.8~m)} &       & \multicolumn{6}{c}{{\bf Condition $\bm{\beta}$2} (1.0~m)} \\
            \cmidrule{3-8}\cmidrule{10-15}\cmidrule{17-22}\cmidrule{24-29}                                                                                                   &                                                  & \bf 1                                    & \bf 2                                            & \bf 3                                   & \bf 4                                          & \bf 5 & \bf 6                                         &  & \bf 1 & \bf 2 & \bf 3 & \bf 4 & \bf 5 & \bf 6 &  & \bf 1 & \bf 2 & \bf 3 & \bf 4 & \bf 5 & \bf 6 &  & \bf 1 & \bf 2 & \bf 3 & \bf 4 & \bf 5 & \bf 6 \\
            \midrule \multirow{2}{*}{\begin{tabular}[c]{c}Method \\ A1\end{tabular}}                                                                                         & \bf 3                                            & 1                                        & -                                                & 2                                       & 1                                              & -     & 1                                             &  & 1     & -     & 4     & -     & -     & -     &  & 2     & 1     & 2     & -     & -     & -     &  & -     & -     & 5     & -     & -     & -     \\
                                                                                                                                                                             & \bf 4                                            & -                                        & 1                                                & -                                       & 3                                              & -     & 1                                             &  & -     & 1     & -     & 4     & -     & -     &  & -     & -     & -     & 5     & -     & -     &  & -     & -     & -     & 5     & -     & -     \\
            \cmidrule(lr){1-2}\cmidrule(lr){3-8}\cmidrule(lr){10-15}\cmidrule(lr){17-22}\cmidrule(lr){24-29} \multirow{2}{*}{\begin{tabular}[c]{c}Method \\ B1\end{tabular}} & \bf 3                                            & -                                        & -                                                & 5                                       & -                                              & -     & -                                             &  & -     & -     & 5     & -     & -     & -     &  & -     & -     & 4     & -     & 1     & -     &  & -     & -     & 5     & -     & -     & -     \\
                                                                                                                                                                             & \bf 4                                            & -                                        & -                                                & -                                       & 5                                              & -     & -                                             &  & -     & -     & -     & 4     & -     & 1     &  & -     & -     & -     & 4     & -     & 1     &  & -     & -     & -     & 4     & -     & 1     \\
            \cmidrule(lr){1-2}\cmidrule(lr){3-8}\cmidrule(lr){10-15}\cmidrule(lr){17-22}\cmidrule(lr){24-29} \multirow{2}{*}{\begin{tabular}[c]{c}Method \\ C1\end{tabular}} & \bf 3                                            & -                                        & -                                                & 5                                       & -                                              & -     & -                                             &  & -     & -     & 5     & -     & -     & -     &  & -     & -     & 4     & -     & 1     & -     &  & -     & -     & 5     & -     & -     & -     \\
                                                                                                                                                                             & \bf 4                                            & -                                        & 1                                                & -                                       & 3                                              & -     & 1                                             &  & -     & -     & -     & 5     & -     & -     &  & -     & -     & -     & 5     & -     & -     &  & -     & -     & -     & 5     & -     & -     \\
            \midrule[0.08em]\addlinespace \multicolumn{2}{c}{\multirow{3}{*}{\bf True Label}}                                                                                & \multicolumn{6}{c}{{\bf Condition $\bm{\gamma}$1} ($45^\circ$)} &                                          & \multicolumn{6}{c}{{\bf Condition $\bm{\gamma}$2} ($90^\circ$)} &                                         & \multicolumn{6}{c}{\bf Condition $\bm{\delta}$}            &       & \multicolumn{6}{c}{\bf Condition $\bm{\varepsilon}$}            \\
            \cmidrule{3-8}\cmidrule{10-15}\cmidrule{17-22}\cmidrule{24-29}                                                                                                   &                                                  & \bf 1                                    & \bf 2                                            & \bf 3                                   & \bf 4                                          & \bf 5 & \bf 6                                         &  & \bf 1 & \bf 2 & \bf 3 & \bf 4 & \bf 5 & \bf 6 &  & \bf 1 & \bf 2 & \bf 3 & \bf 4 & \bf 5 & \bf 6 &  & \bf 1 & \bf 2 & \bf 3 & \bf 4 & \bf 5 & \bf 6 \\
            \midrule \multirow{2}{*}{\begin{tabular}[c]{c}Method \\ A1\end{tabular}}                                                                                         & \bf 3                                            & -                                        & -                                                & 4                                       & -                                              & 1     & -                                             &  & 2     & -     & -     & 1     & 2     & -     &  & -     & -     & 5     & -     & -     & -     &  & -     & 1     & -     & 1     & -     & 3     \\
                                                                                                                                                                             & \bf 4                                            & 3                                        & -                                                & -                                       & 2                                              & -     & -                                             &  & 4     & -     & -     & 1     & -     & -     &  & -     & -     & 2     & 1     & -     & 2     &  & -     & 1     & -     & 3     & -     & 1     \\
            \cmidrule(lr){1-2}\cmidrule(lr){3-8}\cmidrule(lr){10-15}\cmidrule(lr){17-22}\cmidrule(lr){24-29} \multirow{2}{*}{\begin{tabular}[c]{c}Method \\ B1\end{tabular}} & \bf 3                                            & -                                        & -                                                & 5                                       & -                                              & -     & -                                             &  & -     & -     & 5     & -     & -     & -     &  & -     & -     & 4     & -     & 1     & -     &  & -     & -     & 4     & -     & 1     & -     \\
                                                                                                                                                                             & \bf 4                                            & -                                        & -                                                & 3                                       & 2                                              & -     & -                                             &  & -     & -     & 5     & -     & -     & -     &  & -     & 1     & -     & 3     & 1     & -     &  & -     & 2     & 2     & -     & -     & 1     \\
            \cmidrule(lr){1-2}\cmidrule(lr){3-8}\cmidrule(lr){10-15}\cmidrule(lr){17-22}\cmidrule(lr){24-29} \multirow{2}{*}{\begin{tabular}[c]{c}Method \\ C1\end{tabular}} & \bf 3                                            & -                                        & -                                                & 5                                       & -                                              & -     & -                                             &  & -     & -     & 5     & -     & -     & -     &  & -     & -     & 5     & -     & -     & -     &  & -     & -     & 2     & -     & -     & 3     \\
                                                                                                                                                                             & \bf 4                                            & -                                        & 1                                                & 2                                       & 2                                              & -     & -                                             &  & -     & -     & 5     & -     & -     & -     &  & -     & 3     & -     & 2     & -     & -     &  & -     & -     & 3     & -     & -     & 2     \\
            \bottomrule
        \end{tabular}
    \end{table*}
    \subsection{Comparison with Related Works}
    Here we compare the results from this study with those from existing publications,
    as shown in Table~\ref{tab:related_works}. Although these studies assumed different
    measurement conditions (e.g., sensor types, the number of antennas used,
    transmitting power, distance between the antenna and the target person), we can
    see that the proposed method achieves an accuracy that is comparable to that
    reported in the existing studies.

    In particular, the work of Peng~\emph{et al.}~\cite{https://doi.org/10.1109/JSEN.2024.3353256}
    is similar to that of this study because they also used the respiratory and heartbeat
    features for individual identification; they used the respiratory features
    for individual identification, and they also used the heartbeat features only
    during apnea events because the respiratory signals cannot be obtained during
    apnea events. In contrast, this study improves the overall identification accuracy
    by using a combination of the respiratory and heartbeat features simultaneously,
    which differentiates this study from that of Peng \emph{et al.}~\cite{https://doi.org/10.1109/JSEN.2024.3353256}.

    \begin{table}[tb!]
        \centering
        \caption{COMPARISON WITH RELATED WORKS}
        \begin{threeparttable}
            \begin{tabular}{l c c c c c}
                \toprule \multirow{2}{*}{\textbf{Study}}                                        & \multirow{2}{*}{\bf Vital sign} & \multirow{2}{*}{\bf System} & {\bf No. of}   & \multirow{2}{*}{\bf Accuracy} \\
                                                                                                &                                 &                             & {\bf subjects} &                               \\
                \midrule Hwang~\emph{et al.}~\cite{https://doi.org/10.1109/ACCESS.2023.3328641} & HB                              & CW                          & 30             & 88.9\%                        \\
                Peng~\emph{et al.}~\cite{https://doi.org/10.1109/JSEN.2024.3353256}             & Resp. \& HB                     & FMCW                        & 5              & 94.4\%                        \\
                Korany \emph{et al.}~\cite{https://doi.org/10.1109/INFOCOM41043.2020.9155258}   & Resp.                           & Wi-Fi                       & 20             & 95.0\%                        \\
                Wang~\emph{et al.}~\cite{https://doi.org/10.1109/JIOT.2024.3358548}             & Resp.                           & FMCW                        & 37             & 96.0\%                        \\
                Lin~\emph{et al.}~\cite{https://doi.org/10.1145/3117811.3117839}                & HB                              & CW                          & 78             & 98.6\%                        \\
                \cmidrule(lr){1-5} \multirow{2}{*}{\bf This study}                              & \bf Resp. \& HB                 & \bf FMCW                    & \bf 6          & \bf 96.3\%                    \\
                                                                                                & \bf HB                          & \bf CW                      & \bf 30         & \bf 99.4\%                    \\
                \bottomrule
            \end{tabular}
        \end{threeparttable}
        \label{tab:related_works}
    \end{table}

    \section{Conclusion}
    \label{sec:conclusion} In this study, we developed a radar-based individual identification method using both respiratory and heartbeat features. The proposed method extracts the subject's respiratory features using a modified raised-cosine-waveform model. In addition, a parabolic model is used to express the waveform during the expiratory plateau phase. The proposed method extracts the heartbeat
    features by applying a time-frequency analysis and a mel-frequency cepstral analysis to the second time-derivative of the complex radar signal. To identify a suitable combination of features and a classifier, we compared the performances of nine methods based on combinations of three feature vectors, comprising the respiratory feature, the heartbeat feature, and the respiratory and heartbeat feature, and three classifiers, comprising the support vector machine, $k$-nearest neighbors, and multilayer perceptron methods. The accuracy of the proposed method during individual identification was evaluated in two scenarios; one scenario involved five-day radar measurements of six participants conducted by the authors, and the other scenario involved one-day radar measurement of thirty participants, which was provided as a public dataset. Through performance evaluation, we demonstrated that the combination of the respiratory and heartbeat features achieved accuracy of 96.33\% for the five-day radar data from six participants, and accuracy of 99.39\% was realized for a public dataset composed of thirty participants using a short data segment that lasted for only 5.0~s; these results indicate the effectiveness of the proposed approach for radar-based individual identification applications.

    Although our study evaluated the effectiveness of the proposed method using controlled experiments involving six participants (experiment~1) and publicly available data (experiment~2), it is important to note that these experiments were conducted under highly specific and controlled conditions.
    Therefore, there are potential limitations in terms of both the validity of the proposed method and in the general applicability of the findings reported in this study.
    For further evaluation, it will be essential to assess the performance of the proposed method in more realistic scenarios, i.e., where the participants are unaware of the radar measurements and thus behave naturally.
    This will form a critical part of our future research.

    \section*{Acknowledgement}
    \scriptsize
    This work was supported in part by SECOM Science and Technology Foundation, in part by JST under Grants JPMJMI22J2 and JPMJMS2296, in part by JSPS KAKENHI under Grants 21H03427, 23H01420, 23K19119, 23K26115 and 24K17286, and in part by the New Energy and Industrial Technology Development Organization (NEDO). This work involved human subjects in its research. Approval of all ethical and experimental procedures and protocols was granted by the Ethics Committee of the Graduate School of Engineering, Kyoto University, under Approval No. 202223.

    \balance
    \begin{IEEEbiography}
       [{\includegraphics[width=1in,height=1.25in,clip,keepaspectratio]{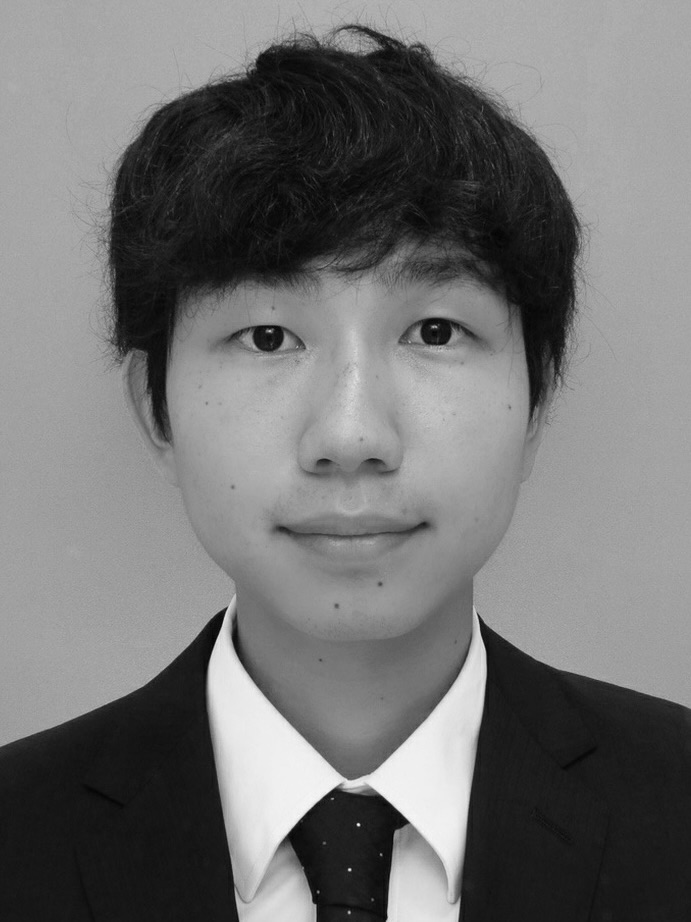}}]{Haruto~Kobayashi}
        (Graduate Student Member, IEEE) received the B.E. degree in electrical
        and electronic engineering from Kyoto University, Kyoto, Japan, in 2023,
        where he is currently pursuing the M.E. degree in electrical engineering
        at the Graduate School of Engineering.
    \end{IEEEbiography}
    \begin{IEEEbiography}
        [{\includegraphics[width=1in,height=1.25in,clip,keepaspectratio]{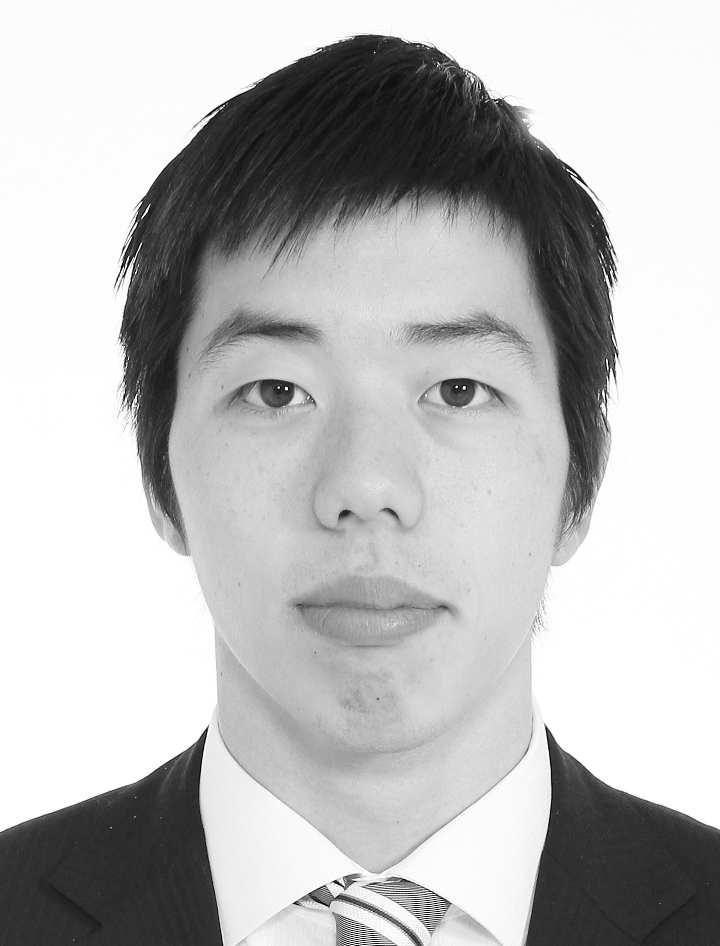}}]{Yuji~Tanaka}
        (Member, IEEE) received the B.E., M.E., and Ph.D. degrees in engineering
        from Kanazawa University, Ishikawa, Japan, in 2015, 2017, and 2023,
        respectively. From 2017 to 2020, he worked at the Information Technology
        R\&D center, Mitsubishi Electric Corporation. He is currently an
        assistant professor at the Nagoya Institute of Technology. His research interests
        include radar signal processing, radio science, and radar measurement of
        physiological signals. He was awarded the Second Prize in the URSI-JRSM
        2022 Student Paper Competition and the Young Researcher's Award from the
        Institute of Electrical and Electronics Engineers (IEICE) Technical
        Committee on Electronics Simulation Technology. He is a member of the IEICE
        and the Society of Geomagnetism and Earth, Planetary and Space Sciences
        (SGEPSS).
    \end{IEEEbiography}
    \begin{IEEEbiography}
        [{\includegraphics[width=1in,height=1.25in,clip,keepaspectratio]{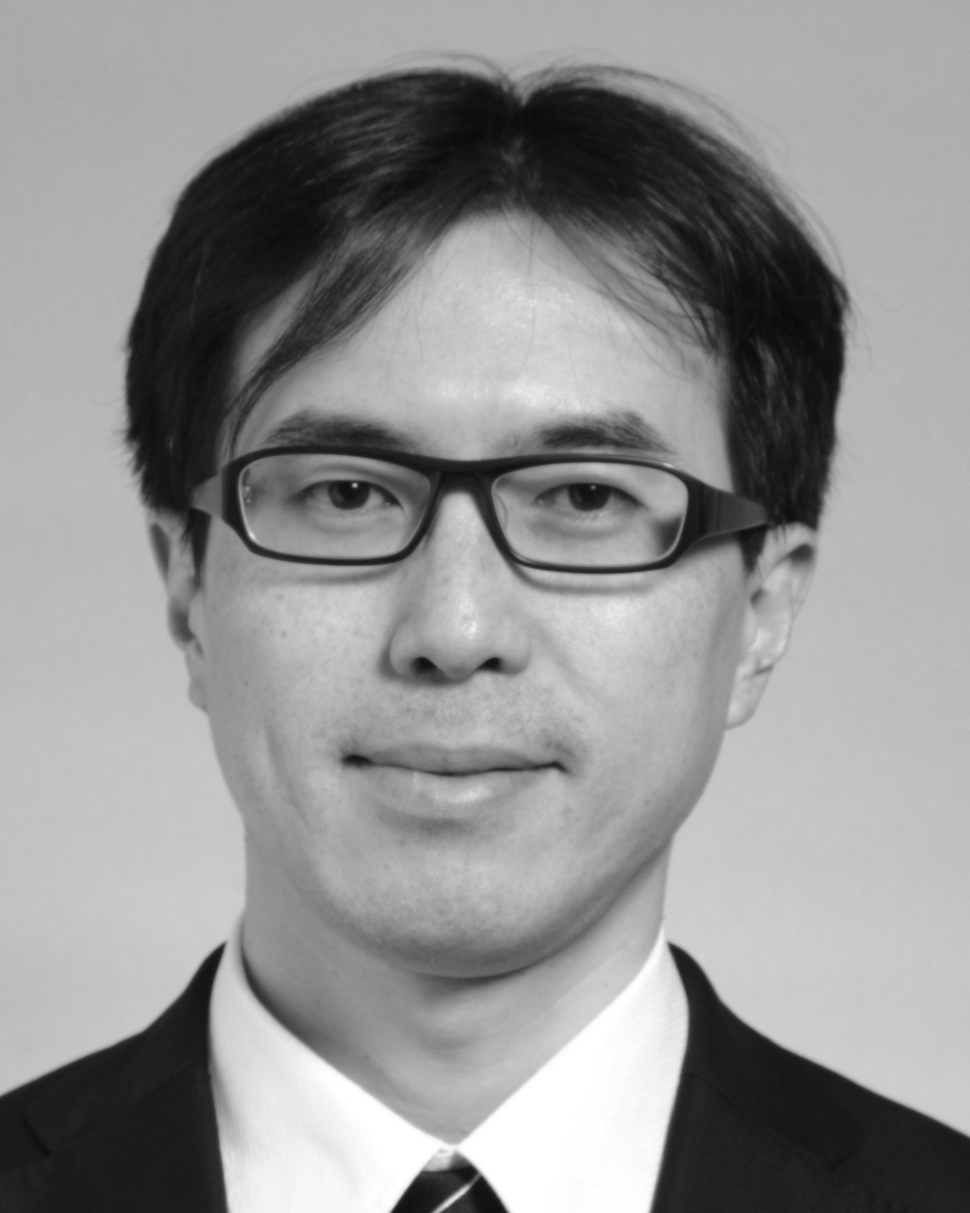}}]
        {Takuya~Sakamoto} (Senior Member, IEEE) received the B.E. degree in electrical
        and electronic engineering from Kyoto University, Kyoto, Japan, in 2000
        and M.I. and Ph.D. degrees in communications and computer engineering
        from the Graduate School of Informatics, Kyoto University, in 2002 and
        2005, respectively. From 2006 through 2015, he was an Assistant Professor
        at the Graduate School of Informatics, Kyoto University. From 2011 through
        2013, he was also a Visiting Researcher at Delft University of
        Technology, Delft, the Netherlands. From 2015 until 2019, he was an
        Associate Professor at the Graduate School of Engineering, University of
        Hyogo, Himeji, Japan. In 2017, he was also a Visiting Scholar at the University
        of Hawaii at Manoa, Honolulu, HI, USA. From 2019 until 2022, he was an
        Associate Professor at the Graduate School of Engineering, Kyoto University.
        From 2018 through 2022, he was also a PRESTO researcher of the Japan Science
        and Technology Agency, Japan. Since 2022, he has been a Professor at the
        Graduate School of Engineering, Kyoto University. His current research
        interests are wireless human sensing, radar signal processing, and radar
        measurement of physiological signals.

        Prof. Sakamoto was a recipient of the Best Paper Award from the International
        Symposium on Antennas and Propagation (ISAP) in 2004, the Young
        Researcher's Award from the Institute of Electronics, Information and Communication
        Engineers of Japan (IEICE) in 2007, the Best Presentation Award from the
        Institute of Electrical Engineers of Japan in 2007, the Best Paper Award
        from the ISAP in 2012, the Achievement Award from the IEICE
        Communications Society in 2015, 2018, and 2023, the Achievement Award from
        the IEICE Electronics Society in 2019, the Masao Horiba Award in 2016, the
        Best Presentation Award from the IEICE Technical Committee on Electronics
        Simulation Technology in 2022, the Telecom System Technology Award from
        the Telecommunications Advancement Foundation in 2022, and the Best Paper
        Award from the IEICE Communication Society in 2007 and 2023.
    \end{IEEEbiography}
\end{document}